\documentclass[showpacs,showkeys,preprintnumbers, amsmath,amssymb,preprint]{revtex4-1}
\usepackage{amssymb}
\usepackage{amsfonts}
\usepackage{amsmath}
\usepackage{graphicx}

\usepackage{graphicx}

\begin{document}

\title{Four-dimensional black holes with scalar hair in nonlinear electrodynamics}
\author{Jos\'{e} Barrientos O.}
\email{josebarrientos@udec.cl}
\affiliation{Departamento de F\'isica, Universidad de Concepci\'on, Casilla 160-C, Concepci\'on, Chile}
\affiliation{Departamento de Ense\~nanza de las Ciencias B\'asicas, Universidad Cat\'olica del Norte, Larrondo 1281, Coquimbo, Chile.}\author{P. A. Gonz\'{a}lez}
\email{pablo.gonzalez@udp.cl}
\affiliation{Facultad de Ingenier\'{\i}a, Universidad Diego Portales, Avenida Ej\'{e}rcito Libertador 441, Casilla 298-V, Santiago, Chile.}
\author{Yerko V\'{a}squez}
\email{yvasquez@userena.cl}
\affiliation{Departamento de F\'{\i}sica y Astronom\'{\i}a, Facultad de Ciencias, Universidad de La Serena,\\
Avenida Cisternas 1200, La Serena, Chile.}
\date{\today}

\begin{abstract}

We consider a gravitating system consisting of a scalar field
minimally coupled to gravity with a self-interacting potential and
a U(1) nonlinear electromagnetic field. Solving analytically and numerically the coupled
system for both power-law and Born-Infeld type electrodynamics, we find charged hairy black
hole solutions.
Then, we study
the thermodynamics of these solutions and we find that at a low temperature the topological charged black hole with scalar hair is thermodynamically preferred, whereas the topological charged black hole without scalar hair is thermodynamically preferred at a high temperature for power-law electrodynamics. Interestingly enough, these phase transitions occur at a fixed critical temperature and do not depend on the exponent $p$ of the nonlinearity electrodynamics.




\end{abstract}

\maketitle


\tableofcontents


\newpage

\section{Introduction}
\label{secs.1}
Hairy black holes are interesting solutions to Einstein's Theory
of Gravity and also to certain types of Modified Gravity Theories. The first attempts to couple a scalar field to gravity was done in an asymptotically flat spacetime finding hairy black hole solutions \cite{BBMB}, but it was realized that these solutions were not physically acceptable as the scalar field was divergent on the horizon and
stability analysis showed that they were unstable \cite{bronnikov}. Also, asymptotically flat black holes with scalar field minimally coupled to gravity were found in \cite{Bechmann:1995sa, Dennhardt:1996cz},  which evade the no hair theorems by allowing partially negative self-interacting potential, which is in conflict with the dominant energy condition. Some of these solutions were found to be stable for some parameter range \cite{Dennhardt:1996cz}.
On the other hand, by introducing a cosmological constant,  hairy black hole solutions with a minimally
coupled scalar field and a self-interaction potential in
asymptotically dS space were found, but unstable \cite{Zloshchastiev:2004ny,Torii:1998ir}. Also, a hairy
black hole configuration was reported  for scalar field non-minimally coupled to gravity \cite{Martinez:2002ru}, but
a perturbation analysis showed the instability of the solution
\cite{Harper:2003wt,Dotti:2007cp}.
 In the case of a
negative cosmological constant, stable solutions were found
numerically for spherical geometries \cite{Torii:2001pg,
Winstanley:2002jt} and an exact solution in asymptotically AdS space with hyperbolic geometry was presented in \cite{Martinez:2004nb}. A study of general properties of black holes with scalar hair with spherical symmetry can be found in Ref. \cite{Bronnikov:2001tv}.
Further hairy solutions 
were reported in
\cite{Anabalon:2012ta,Anabalon:2012ih, Anabalon:2012tu,Bardoux:2012tr, Gonzalez:2013aca, Kolyvaris:2011fk, Kolyvaris:2013zfa, Anabalon:2013oea, Cisterna:2014nua}
with various properties. 
Furthermore, charged hairy solutions were also found, for instance
in \cite{Martinez:2005di}, a topological black
hole dressed with a conformally coupled scalar field and
  electric charge was studied. 
 An electrically charged black hole solution with a scalar field minimally coupled to gravity and electromagnetism
 was presented in \cite{Martinez:2006an}. Recently, for a gravitating system consisting of a scalar field
minimally coupled to gravity with a self-interacting potential and
 U(1) electromagnetic field, exact charged hairy black
hole solutions with the scalar field which is regular outside the event horizon have been found in \cite{Gonzalez:2014tga, Xu:2014uka, Fan:2015oca}.
Also, new hairy black hole solutions, boson
stars and numerical rotating hairy black hole solutions were
discussed
\cite{Dias:2011at,Stotyn:2011ns,Dias:2011tj,Kleihaus:2013tba,Herdeiro:2014goa,Herdeiro:2014ima}, as well as time-dependent hairy black holes \cite{Zhang:2014sta, Lu:2014eta}. For a review of hairy black holes we refer the reader to \cite{Volkov:2016ehx}.



In this work, we extend our previous work \cite{Gonzalez:2013aca, Gonzalez:2014tga} and we consider a gravitating system consisting of a scalar field
minimally coupled to gravity with a self-interacting potential and
U(1) nonlinear electromagnetic field. Then, we obtain black hole solutions for power-law and Born-Infeld type electrodynamics and we study the thermodynamics and the phase transitions between hairy charged black holes dressed with a scalar field and no hairy charged black holes, focusing on the effects of the nonlinearity of the Maxwell source. The interest in nonlinear electrodynamics arises 
with the order to eliminate the problem of infinite energy of the electron by  Born and Infeld \cite{BI}. 
Also, nonlinear electrodynamics
emerge in the modern context of the low-energy limit of heterotic string theory \cite{Kats:2006xp,Anninos:2008sj,Cai:2008ph},
and  play an important role 
in the construction of regular black hole solutions \cite{AyonBeato:1998ub, AyonBeato:1999rg, Cataldo:2000ns, Bronnikov:2000vy, Burinskii:2002pz, Matyjasek:2004gh}.
Some black holes/branes solutions in a nonlinear electromagnetic field have been investigated for instance in \cite{Cai:2004eh, Dey:2004yt, Aiello:2004rz, Hendi:2009sw, Hendi:2010zz, Hendi:2013dwa, Hendi:2015hoa, Hendi:2016dmh} and references therein. The thermodynamics of Einstein-Born-Infeld black holes with a negative cosmological constant was studied in \cite{Miskovic:2008ck} and for a power-law electrodynamic in \cite{Gonzalez:2009nn}, where the authors  showed that a set of small black holes are locally stable by computing the heat capacity and the electrical permittivity. The thermodynamics of Gauss-Bonnet black holes for a power-law electrodynamic was studied in \cite{Hendi:2010zza}. On the other hand, higher dimensional black hole solutions to Einstein-dilaton theory coupled to Maxwell field were found in \cite{Sheykhi:2009pf, Hendi:2015xya} and black hole solutions to Einstein-dilaton theory coupled to Born-Infeld  and power-law electrodynamics were found in \cite{Dehghani:2006zi, Zangeneh:2015wia}.

The phase transitions have been of great interest since the discovery of a phase transition by Hawking and Page in a four-dimensional Schwarzschild AdS background \cite{HP}. Witten \cite{Witten:1998zw} has extended this four-dimensional transition to arbitrary dimension and provided a natural explanation of a confinement/deconfinement transition on the boundary field theory via the AdS/CFT correspondence. However, phase transitions have recently garnered a great deal of attention motivated mainly by the relationship between the phase transitions and holographic superconductivity \cite{Gubser:2008px, Hartnoll:2008vx} in the context of the AdS/CFT correspondence. Furthermore, the effects of nonlinear electrodynamics on the properties of the holographic superconductors have recently been investigated \cite{Jing:2011vz, Zhao:2012cn, Roychowdhury:2012hp,  Gangopadhyay:2013qza, Dey:2014xxa, Dey:2014voa, Lai:2015rva, Ghorai:2015wft, Liu:2015lit, Sheykhi:2016kqh}. It is known that these phase transitions can be obtained by considering black holes as states in a same grand canonical ensemble and by comparing the free energy associated with each of them. So, it is necessary to find the conserved charge of the theory. It is worth mentioning that the phase transition phenomena have been analyzed and  classified by exploiting Ehrenfest's scheme \cite{Banerjee:2010da, Banerjee:2010bx,  Banerjee:2011au, Banerjee:2011raa}. Another point of view to study  phase transitions is to consider Bragg-Williams' construction of a free energy function \cite{Banerjee:2010ve}. 
Also, it was shown that if the space is flat, then the Reissner-Nordstr\"om black hole is thermodynamically preferred, whereas if the space is AdS the hairy charged black hole is thermodynamically preferred at a low temperature \cite{Gonzalez:2014tga}.

 The work is organized as follows. In Section \ref{secs.2} we
 present the general formalism. Then, we derive  the field equations and we find hairy black hole
 solutions. In Section \ref{secs.4}
 we study the thermodynamics of
 our solutions and in Section \ref{secs.5} we present our conclusions.

\section{Four-dimensional black holes with scalar hair in nonlinear electrodynamics}
\label{secs.2}

The  four-dimensional Einstein-Hilbert action with a scalar field minimally coupled to
curvature having a self-interacting potential $ V(\phi)$ in the
presence of a nonlinear electromagnetic field is
 \begin{eqnarray} \label{action}
 I=\int d^{4}x\sqrt{-g}\left(\frac{1}{2 \kappa }R+\mathcal{L}(F^2)
 -\frac{1}{2}g^{\mu\nu}\nabla_{\mu}\phi\nabla_{\nu}\phi-V(\phi)\right)~,
 \end{eqnarray}
where $\kappa=8 \pi G$, with $G$ the Newton constant and $\mathcal{L}(F^2)$ an arbitrary function of the electromagnetic invariant $F^2=F_{\alpha \beta} F^{\alpha \beta}$
. The
resulting field equations from the above action are
 \begin{eqnarray}
 R_{\mu\nu}-\frac{1}{2}g_{\mu\nu}R=\kappa (T^{(\phi)}_{\mu\nu}+T^{(F)}_{\mu\nu})~,\label{field1}
 \end{eqnarray}
where the energy-momentum tensors $T^{(\phi)}_{\mu\nu}$ and $T^{(F)}_{\mu\nu}$ for the
scalar and electromagnetic fields are
 \begin{eqnarray}
 \nonumber T^{(\phi)}_{\mu\nu}&=&\nabla_{\mu}\phi\nabla_{\nu}\phi-
 g_{\mu\nu}\left[\frac{1}{2}g^{\rho\sigma}\nabla_{\rho}\phi\nabla_{\sigma}\phi+V(\phi)\right]~,\\
T^{(F)}_{\mu\nu}&=&g_{\mu \nu} \mathcal{L}(F^2)-4 \frac{d \mathcal{L}(F^2)}{dF^2} F_{\mu} ^{\,\,\, \lambda} F_{\nu \lambda}~,  
 \label{energymomentum}
 \end{eqnarray}
  respectively. Using Eqs. (\ref{field1}) and (\ref{energymomentum}) we obtain the equivalent equation
 \begin{eqnarray}
R_{\mu\nu}-\kappa\left(\partial_\mu\phi \partial_\nu\phi+g_{\mu\nu}V(\phi)\right)=\kappa \left(-4 \frac{d \mathcal{L}(F^2)}{dF^2}F_{\mu}^{\,\,\, \alpha}F_{\nu \alpha} +2 g_{\mu \nu}F^2 \frac{d \mathcal{L}(F^2)}{dF^2} - g_{\mu \nu} \mathcal{L}(F^2)\right)~. \label{einstein1}
 \end{eqnarray}
Now, if we consider the following metric ansatz
 \begin{eqnarray}
 ds^{2}=-f(r)dt^{2}+f^{-1}(r)dr^{2}+a^{2}(r)d \Omega^2 ~,\label{metricBH}
 \end{eqnarray}
where $d \Omega ^2$ is the metric of the spatial 2 section, which
can have a positive, negative or zero  curvature, and
$A_{\mu}=(A_t(r),0,0,0)$ is the scalar potential of the
electromagnetic field. We find the
following three independent differential equations 
\begin{equation}\label{first}
f^{\prime\prime}(r)+2\frac{a^{\prime}(r)}{a(r)}f^{\prime}(r)+2 \kappa V(\phi)=2 \kappa \mathcal{L}(F^2)~,
\end{equation} 
\begin{equation} 
f^{\prime\prime}(r)+2\frac{a^{\prime}(r)}{a(r)}f^{\prime}(r)+\left(4\frac{a^{\prime\prime}(r)}{a(r)}+2\kappa \phi^{\prime}(r)^{2}\right)f(r)+2\kappa V(\phi)=2 \kappa \mathcal{L}(F^2)~,
\end{equation}
\begin{equation}
\frac{a^{\prime}(r)}{a(r)}f^{\prime}(r)+\left(\left(\frac{a^{\prime}(r)}{a(r)}\right)^{2}+\frac{a^{\prime\prime}(r)}{a(r)}\right)f(r)-\frac{k}{a(r)^{2}}+\kappa V(\phi)= \kappa \mathcal{L}(F^2)+4\kappa A_t^{\prime} (r)^2 \frac{d \mathcal{L}(F^2)}{dF^2}~,
\end{equation}
where $k=1,0,-1$ parameterizes the curvature of the spatial 2-section.  So, if we eliminate the potential $V(\phi)$ from the above equations we obtain
\begin{equation}
a^{\prime\prime}(r)+\frac{1}{2}\kappa \phi^{\prime}(r)^{2}a(r)=0\label{adiff}~,
\end{equation}
\begin{equation}\label{parafprimaprima}
f^{\prime\prime}-2\left(\left(\frac{a^{\prime}(r)}{a(r)}\right)^{2}+\frac{a^{\prime\prime}(r)}{a(r)}\right)f(r)+\frac{2k}{a(r)^{2}}=-8 \kappa A_t^{\prime}(r)^2 \frac{d \mathcal{L}(F^2)}{dF^2} ~.
\end{equation}
In the following, we will work in units where $\kappa=1$. By considering the scalar field studied in \cite{Gonzalez:2014tga}, 
\begin{equation}
\phi \left( r\right) =\frac{1}{\sqrt{2}}\ln \left( 1+\frac{\nu }{r}\right) ~,
\label{field}
\end{equation}
where $\nu $ is a parameter controlling the behavior of the
scalar field and it has the dimension of length, from
Eqs. (\ref{adiff}) and (\ref{field}) we determine the
function $a(r)$, which reads
\begin{equation}\label{eqa}
a\left( r\right) =\sqrt{r\left( r+\nu \right) }~.
\end{equation}
Also, from the Maxwell equations
\begin{equation}\label{gmaxw}
\partial_{\mu}\left(\sqrt{-g}F^{\mu\nu} \frac{d \mathcal{L}(F^2)}{dF^2} \right)=0~, 
\end{equation}
we obtain the following relation
\begin{equation}\label{maxw}
A_{t}^{\prime}(r) \frac{d \mathcal{L}(F^2)}{dF^2}=-\frac{\tilde{Q}}{a(r)^2}~, 
\end{equation}
where $\tilde{Q}$ is an integration constant. We can also
determine the metric function $f(r)$ replacing \eqref{eqa} and \eqref{maxw} in \eqref{parafprimaprima}
\begin{equation}\label{metricfunction}
\begin{aligned}
f(r)&=-\frac{\Lambda}{3}r^2-\frac{1}{3}\nu(6\alpha_{2}+\Lambda)r+k-\alpha_{2} \nu^2-2\alpha_{2}r(r+\nu) \ln \left(\frac{r}{r+\nu}\right) \\
&-8r(r+\nu)\int\frac{\int r (r+ \nu)A_t^{\prime} (r)^2 \frac{d \mathcal{L}(F^2)}{dF^2} dr}{r^2(r+\nu)^2}dr \\
&=-\frac{\Lambda}{3}r^2-\frac{1}{3}\nu(6\alpha_{2}+\Lambda)r+k-\alpha_{2} \nu^2-2\alpha_{2}r(r+\nu) \ln \left(\frac{r}{r+\nu}\right) \\
&\quad +8\tilde{Q} r(r+\nu)\int_{\infty}^r\frac{A_{t}(r)}{r^2(r+\nu)^2}dr~.
\end{aligned}
\end{equation}
To find hairy black hole solutions the differential equations
have to be supplemented with the Klein-Gordon equation of the scalar field, which in general coordinates reads
\begin{equation}\label{klg}
\nonumber  \Box \phi =\frac{d V}{d \phi}~,
\end{equation}
whose solution for the self-interacting potential $ V(r)$ is
\begin{equation}\
V(r)=\frac{\nu^2}{2}\int \frac{f^{\prime}(r)}{r^2(r+\nu)^2}dr~.
\end{equation}
In the following sections we will consider two specific electromagnetic Lagrangians $\mathcal{L}(F^2)$: Power-law and Born-Infeld type electrodynamics.

\subsection{Power-law electrodynamics}
In this section we consider power-law electrodynamics characterized by the following Lagrangians  \cite{Gurtug:2010dr}
\begin{equation}\label{powerlaw}
\mathcal{L}(F^2)=\eta | F^2| ^p~,
\end{equation}
where $p$ is a rational number and the absolute value ensures that any configuration of electric and magnetic fields can be described by these Lagrangians. One could also consider the Lagrangian without the absolute value and the exponent $p$ restricted to being an integer or a rational number with an odd denominator \cite{Hassaine:2008pw}. The sign of the coupling constant $\eta$ will be chosen such that the energy density of the electromagnetic field is positive; that is, the minus $T^{t\,(F)}_{\,\, t}$ component of the electromagnetic energy-momentum tensor must be positive
\begin{equation}
-T^{t\,(F)}_{\,\, t}=\eta |F^2|^p(2p-1) >0~.
\end{equation}
This condition is guaranteed in the following cases: $p>1/2$ and $\eta>0$ or $p<1/2$ and $\eta<0$.  In what follows we focus our attention on purely electric field configurations.

\begin{figure}[h]
\begin{center}
\includegraphics[width=0.4\textwidth]{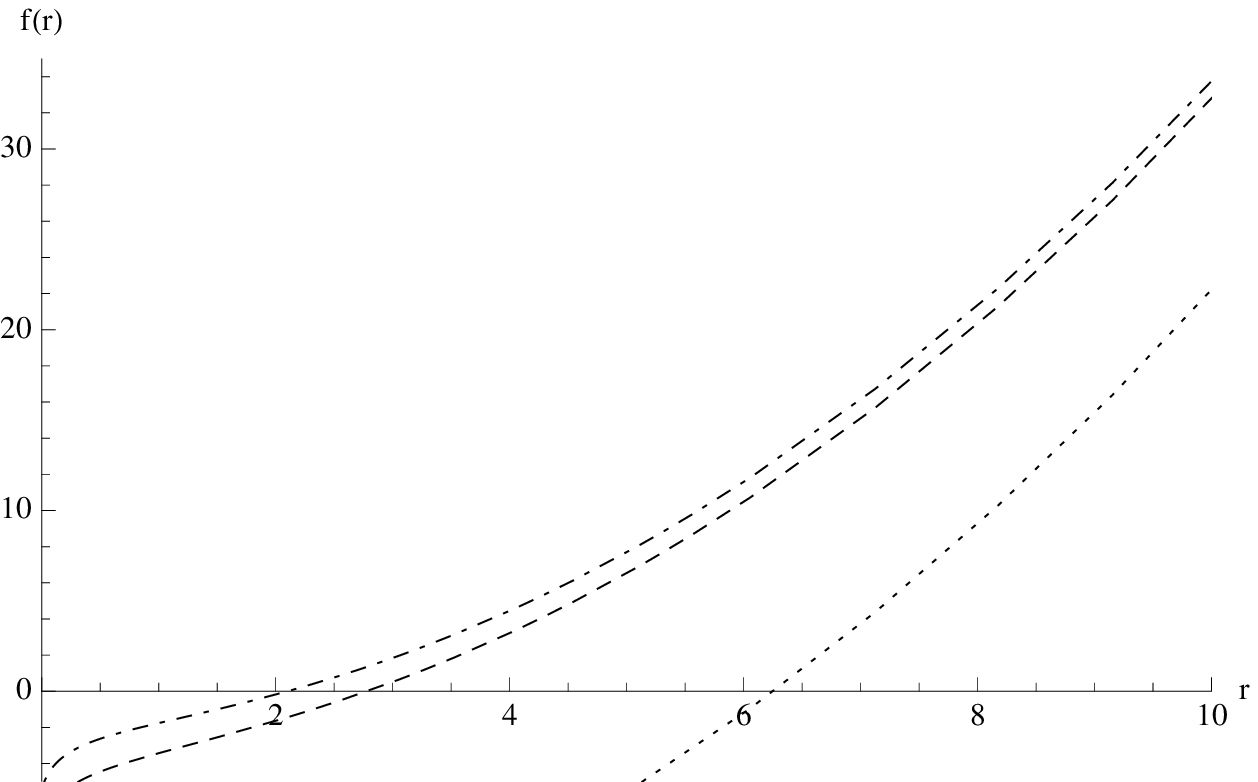}
\includegraphics[width=0.4\textwidth]{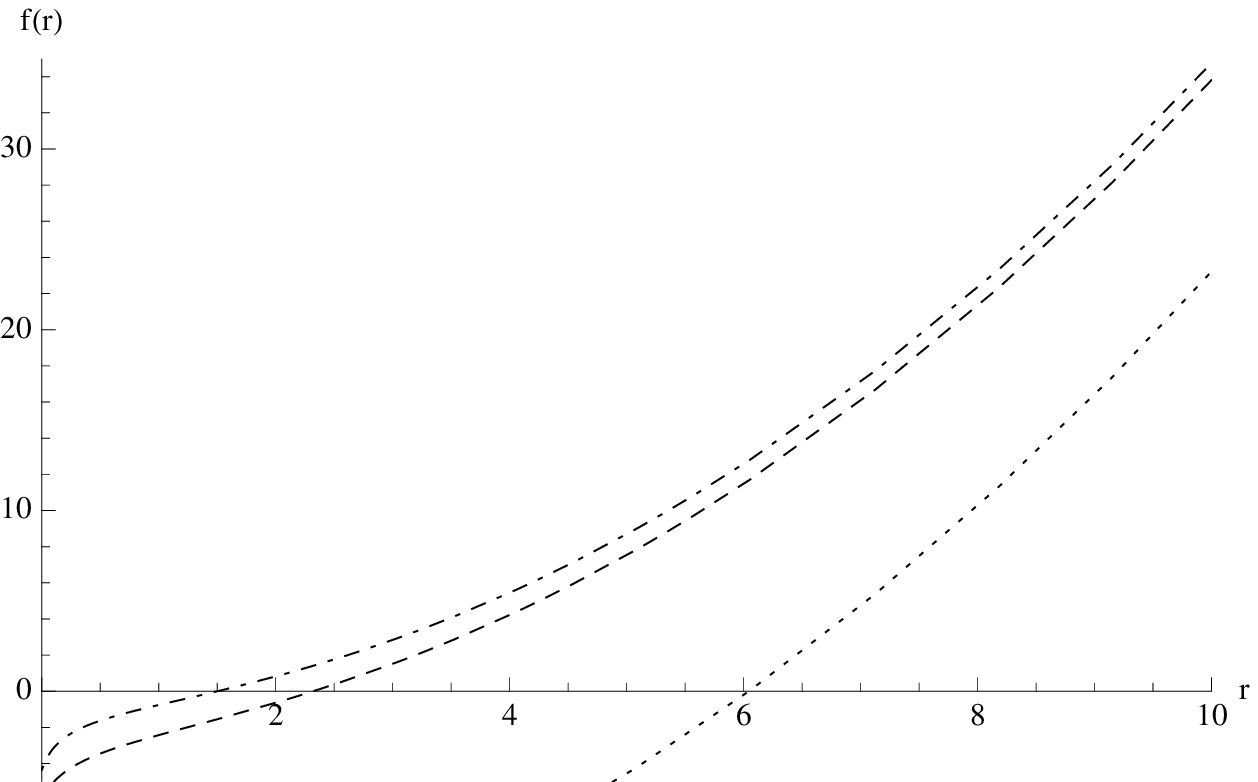}
\includegraphics[width=0.4\textwidth]{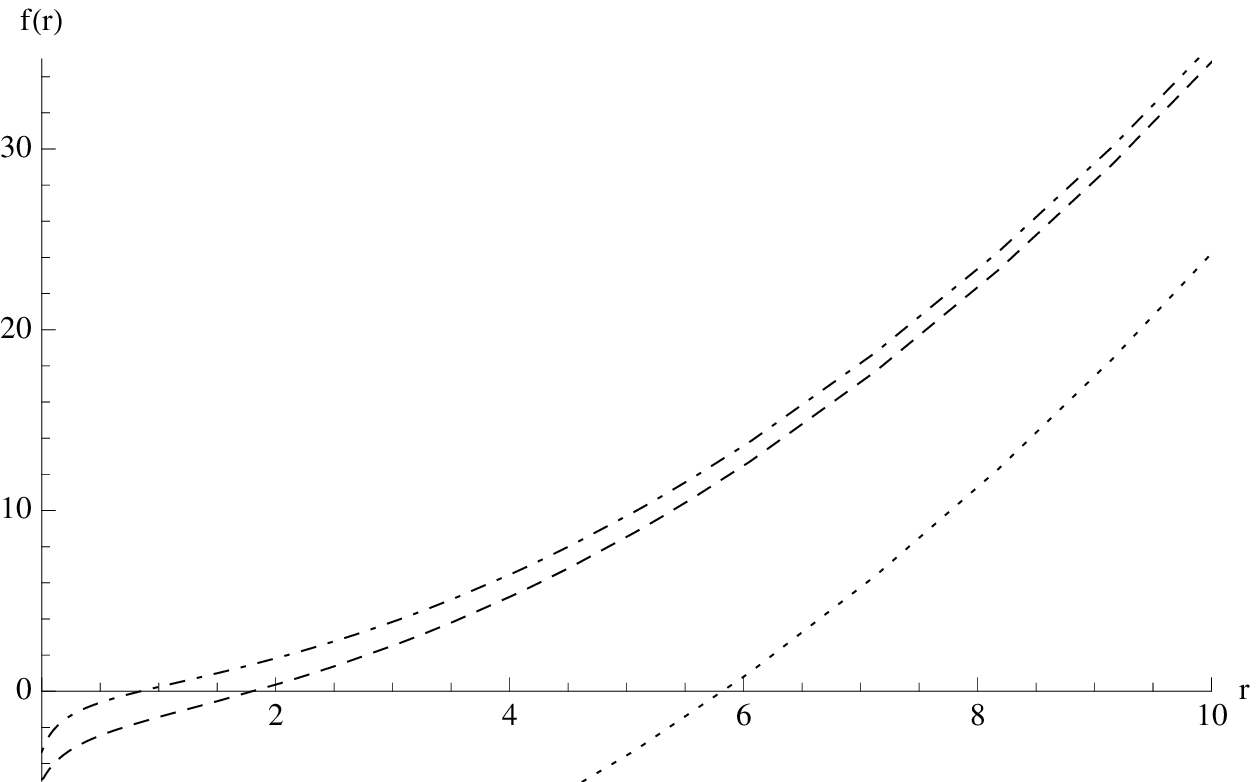}
\end{center}
\caption{The behavior of $f(r)$ for $\nu =0.5$, $\Lambda=-1$, $\alpha_1=1$, $\alpha_2=1$, $\eta=1/4$, $p=5$ (dotted line), $p=3$ (dashed line) and $p=2$ (dot-dashed line). Left figure for $k=-1$, right figure for $k=0$, and bottom figure for $k=1$.}
\label{plots0}
\end{figure}

\begin{figure}[h]
\begin{center}
\includegraphics[width=0.4\textwidth]{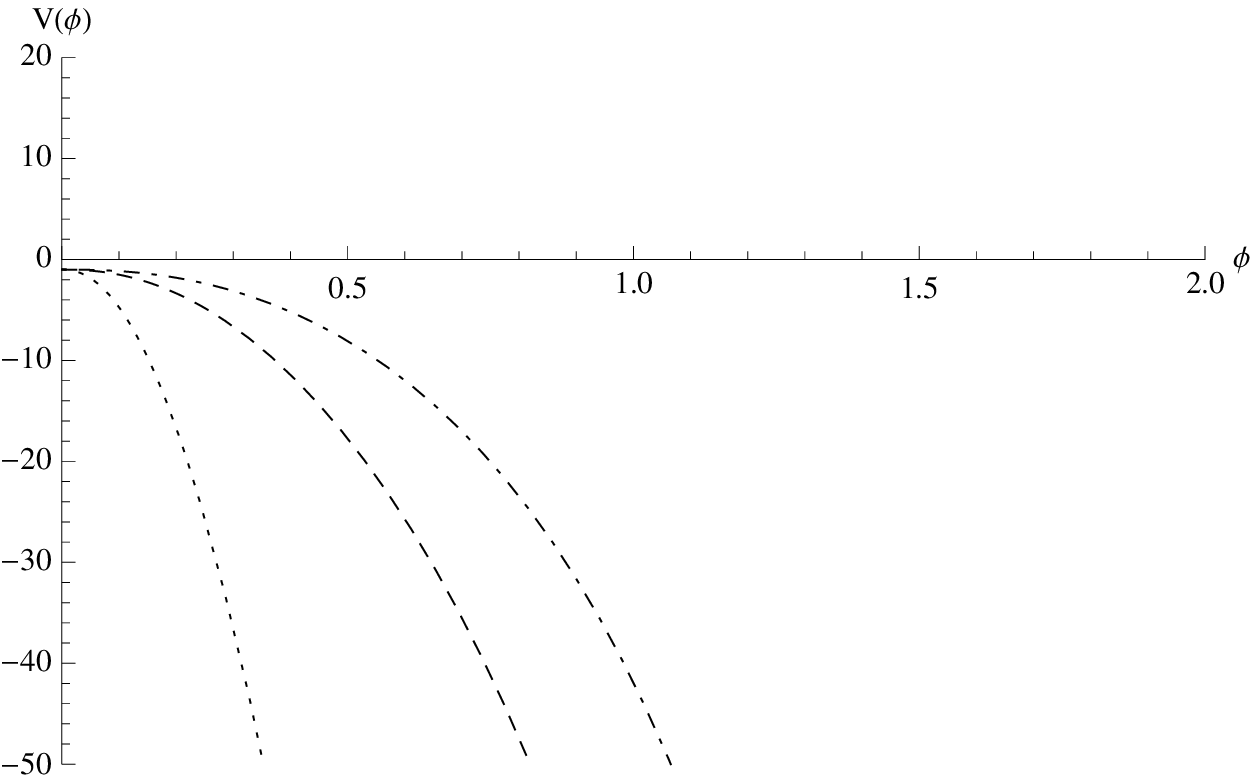}
\end{center}
\caption{The behavior of $V(\phi)$ for $\Lambda=-1$, $\alpha_1=1$, $\alpha_2=1$, $\eta=1/4$, $p=5$ (dotted line), $p=3$ (dashed line) and $p=2$ (dot-dashed line). 
} \label{plots11}
\end{figure}

\begin{figure}[h]
\begin{center}
\includegraphics[width=0.4\textwidth]{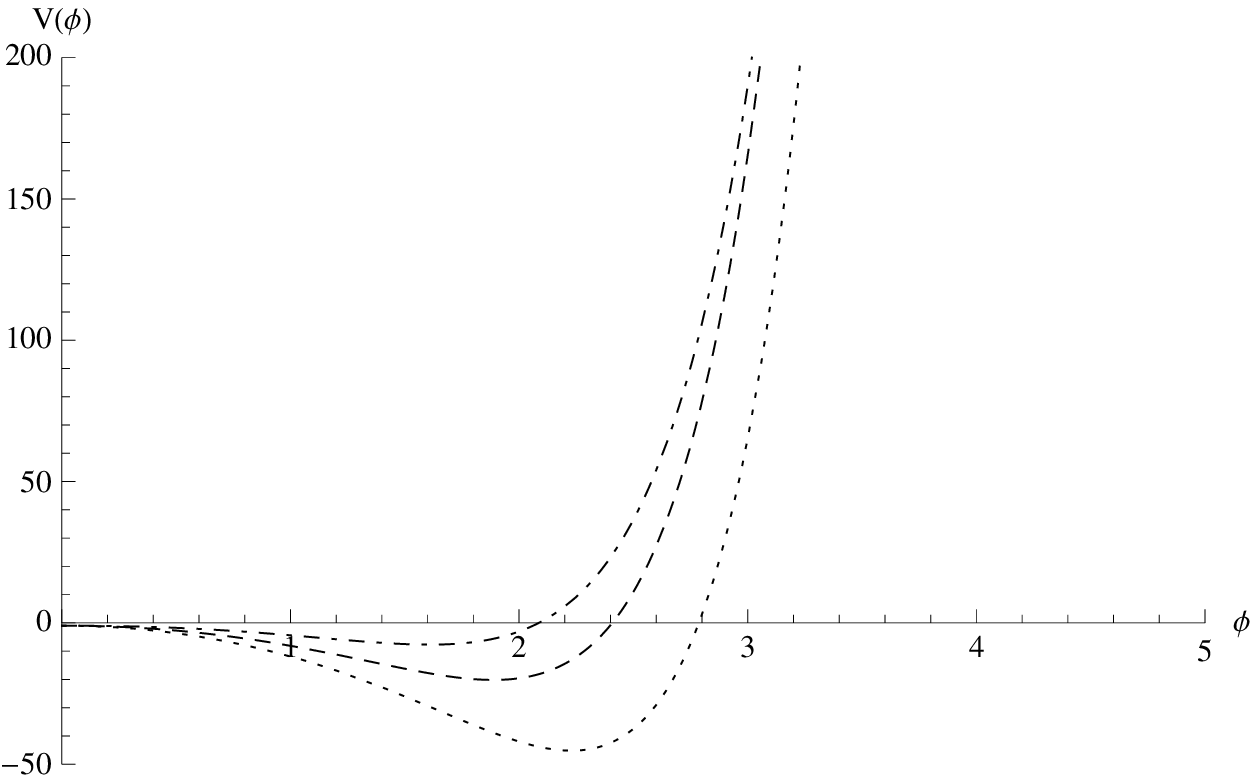}
\end{center}
\caption{The behavior of $V(\phi)$ for $\Lambda=-1$, $\alpha_1=0.1$, $\alpha_2=-5$, $\eta=1/4$, $p=3.5$ (dotted line), $p=3$ (dashed line) and $p=2$ (dot-dashed line). 
} \label{plott10}
\end{figure}

\begin{figure}[h]
\begin{center}
\includegraphics[width=0.4\textwidth]{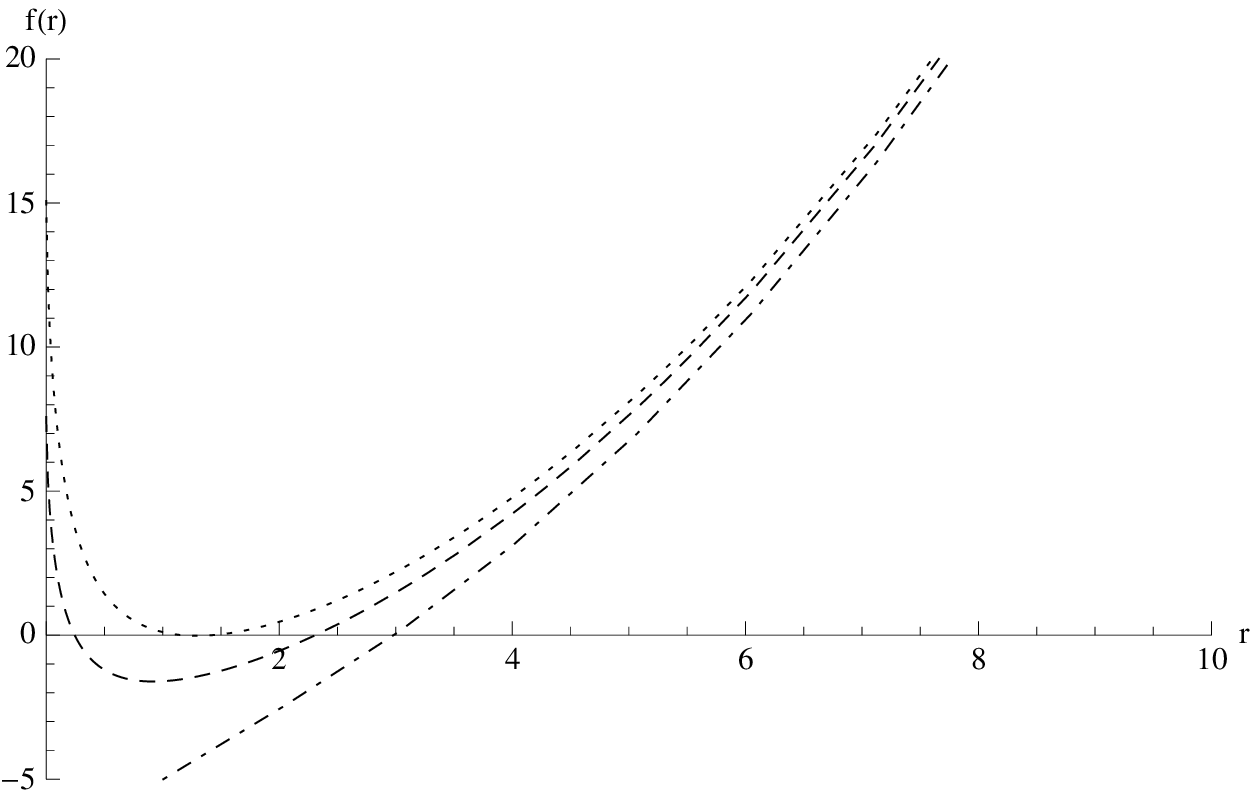}
\includegraphics[width=0.4\textwidth]{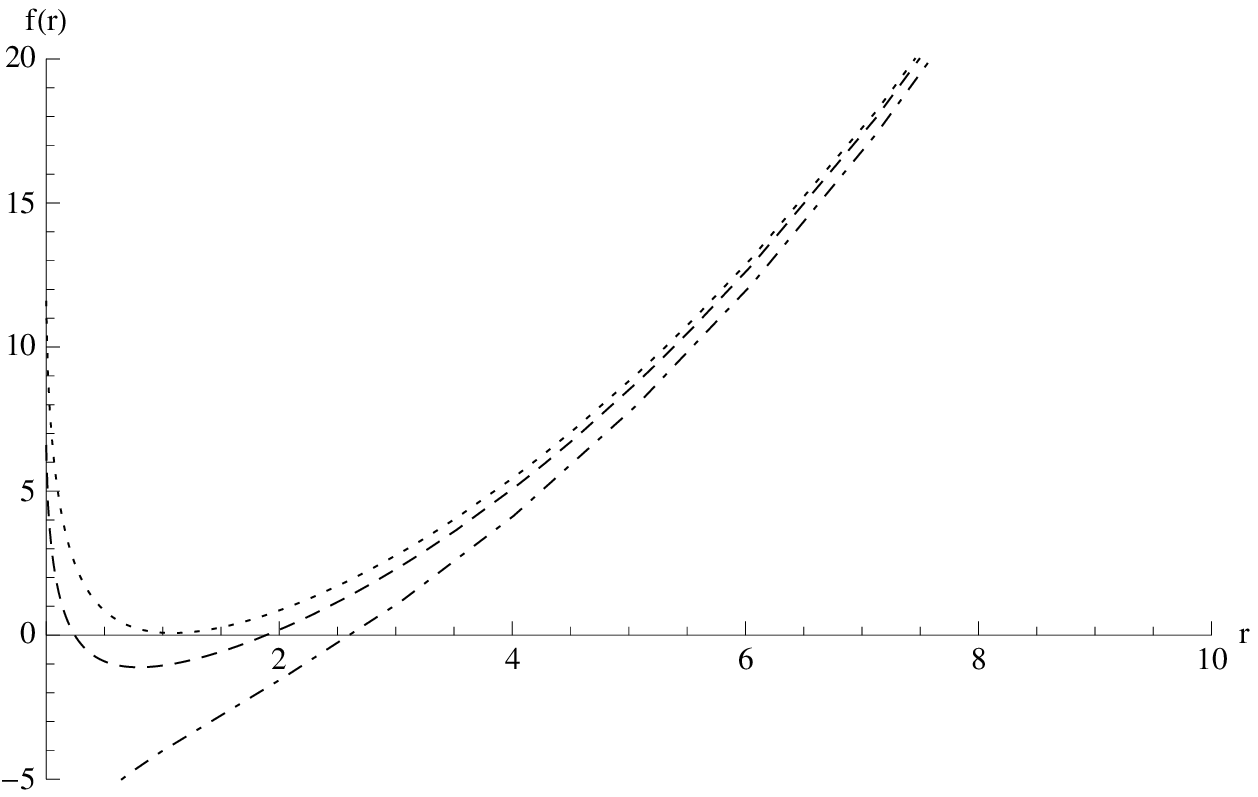}
\includegraphics[width=0.4\textwidth]{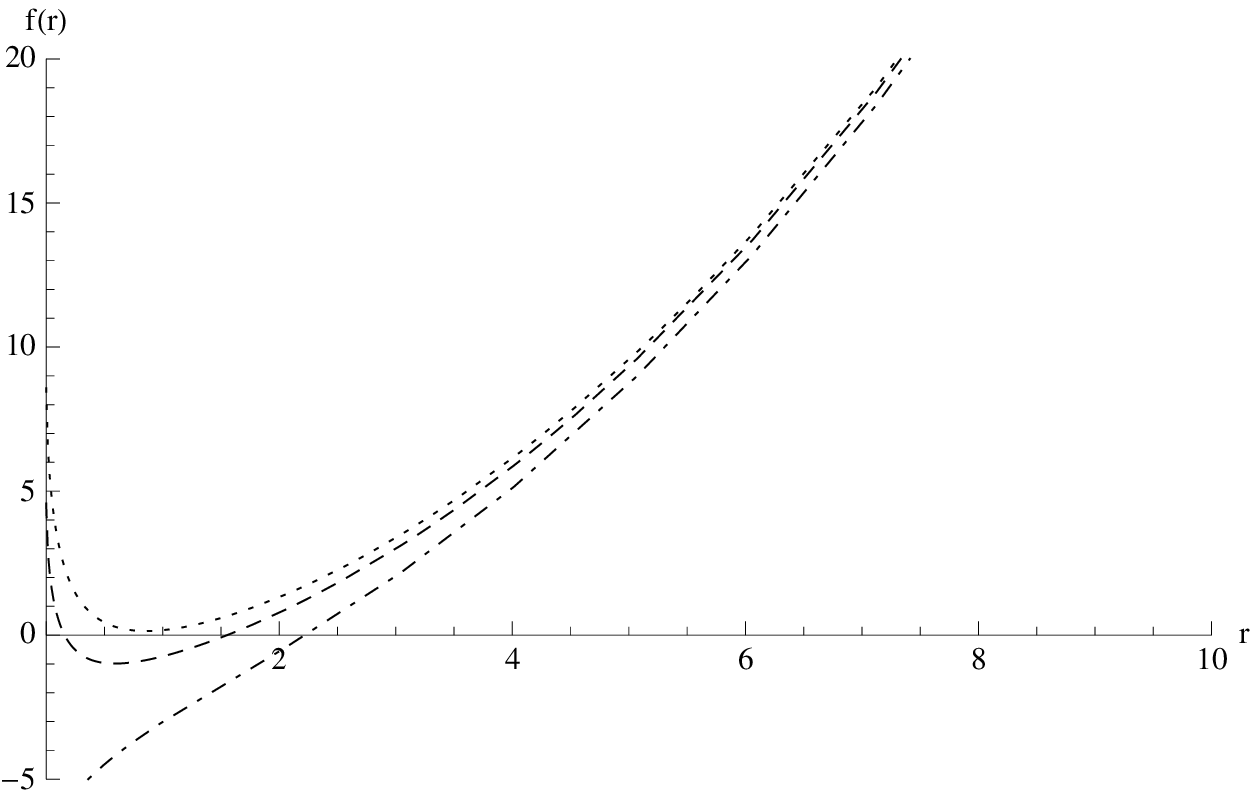}
\end{center}
\caption{The behavior of $f(r)$ for $\nu =1$, $\Lambda=-1$, $\alpha_1=1$, $\eta=1/4$, $p=2$. Left figure for $k=-1$ and $\alpha_2=-10$ (dot-dashed line), $\alpha_2=-25$ (dashed line), $\alpha_2=-32.5$ (dotted line). Right figure for $k=0$ and $\alpha_2=-10$ (dot-dashed line), $\alpha_2=-23$ (dashed line), $\alpha_2=-28$ (dotted line). Bottom figure for $k=1$ and $\alpha_2=-10$ (dot-dashed line), $\alpha_2=-20$ (dashed line), $\alpha_2=-24$ (dotted line).}
\label{plots00}
\end{figure}

\begin{figure}[h]
\begin{center}
\includegraphics[width=0.4\textwidth]{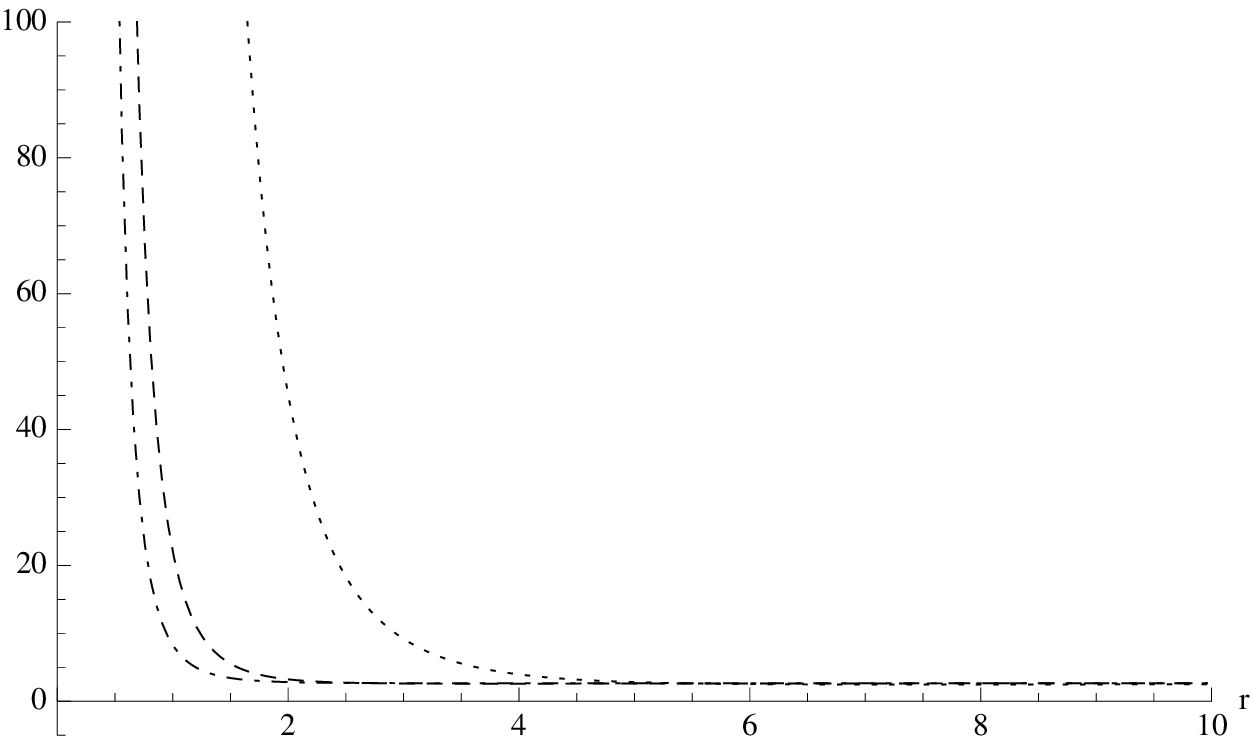}
\includegraphics[width=0.4\textwidth]{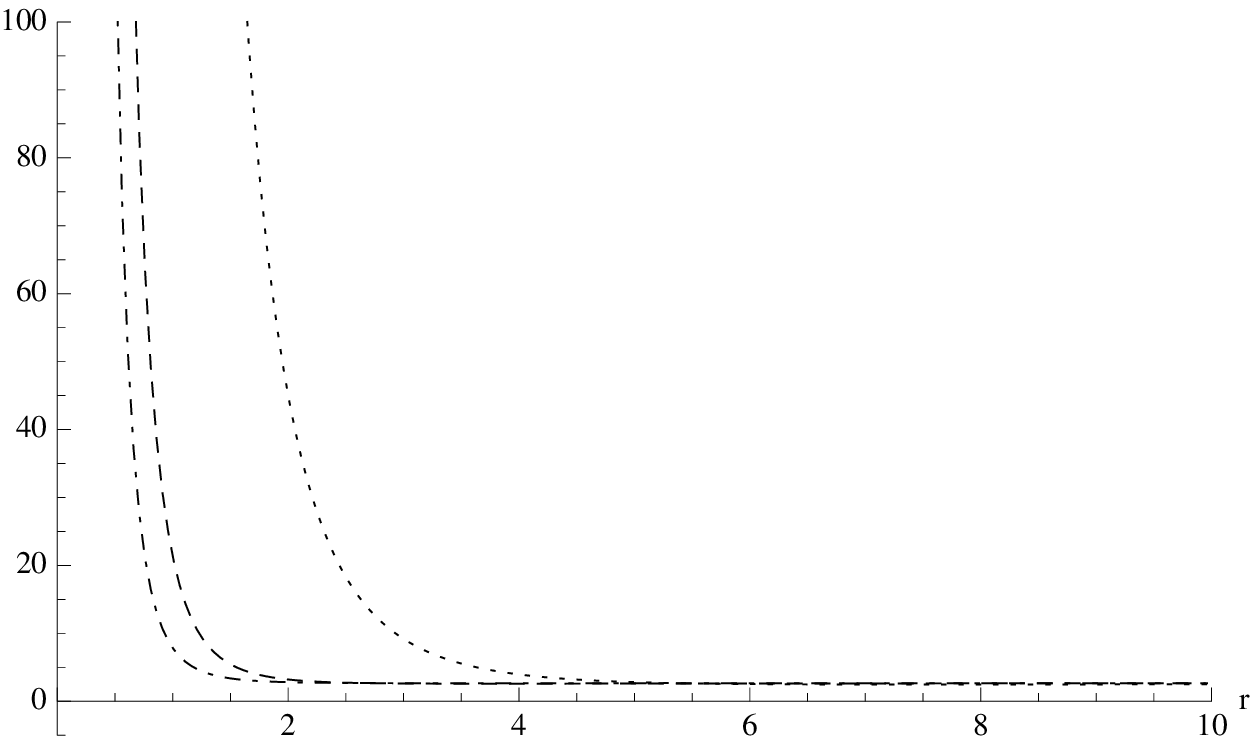}
\includegraphics[width=0.4\textwidth]{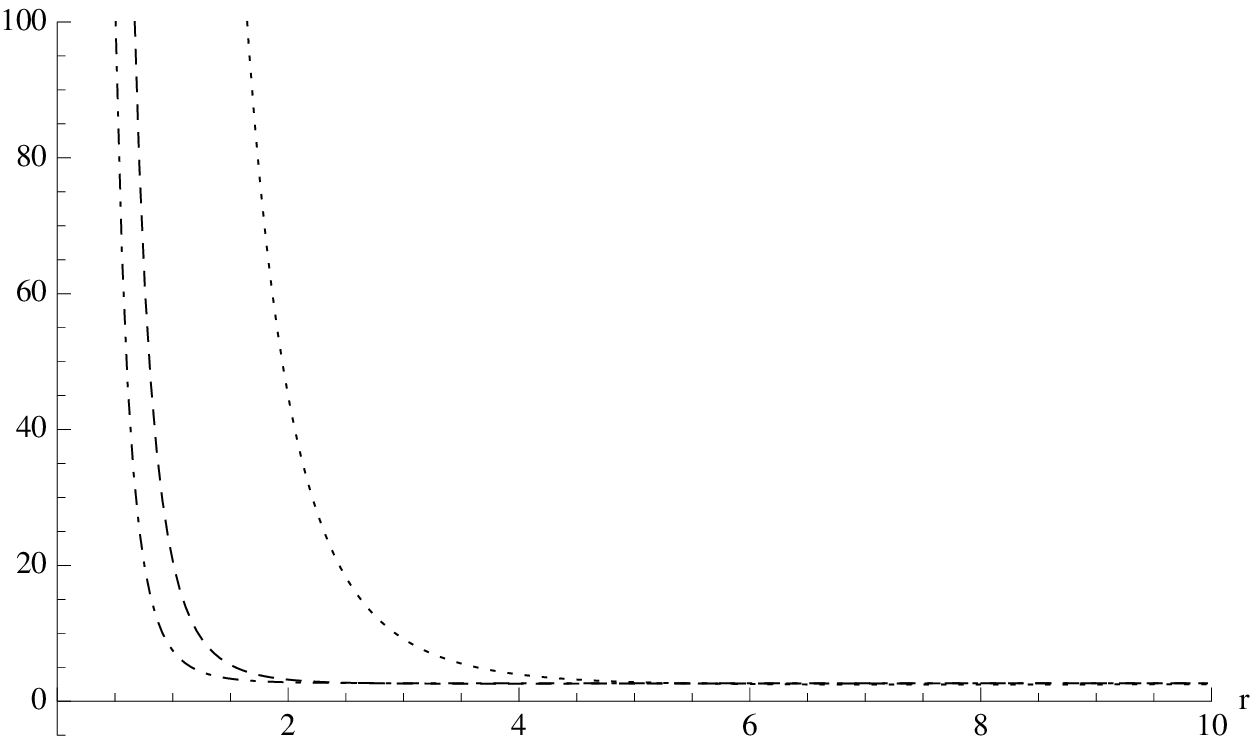}
\end{center}
\caption{The behavior of Kretschmann scalar $R_{\mu\nu\rho\sigma}R^{\mu\nu\rho\sigma}(r)$ as a function of $r$  for $\nu =0.5$, $\Lambda=-1$, $\alpha_1=1$, $\alpha_2=1$, $\eta=1/4$, $p=5$ (dotted line), $p=3$ (dashed line) and $p=2$ (dot-dashed line). Left figure for $k=-1$, right figure for $k=0$, and bottom figure for $k=1$.} \label{figuraRR}
\end{figure}

From Maxwell equations (\ref{maxw}) we obtain 
\begin{equation}
A_{t}^{\prime}(r)=\frac{Q}{(r(r+\nu))^{1/(2p-1)}}~,
\end{equation} 
where $Q$ is an integration constant related to $\tilde{Q}$ by $\tilde{Q}=\eta 2^{p-1} p (Q^{2})^{ p}/Q$. So
\begin{equation}\label{eqA}
A_{t}(r)=\frac{(2p-1)Qr^{(2-2p)/(1-2p)} (1/ \nu)^{1/(-1+2p)}}{2(p-1)}\,_2F_1\left[\frac{1}{2p-1},\,\frac{2(p-1)}{2p-1}~,\,\frac{4p-3}{2p-1},\,-\frac{r}{\nu}\right]+C~.
\end{equation}
The integration constant $C$ will be chosen in such a way the electric potential goes to zero asymptotically for $1/2<p<3/2$, and is given by
\begin{equation}
C=-Q (1/ \nu)^{\frac{3-2p}{-1+2p}} \frac{\Gamma \left( \frac{2-2p}{1-2p} \right) \Gamma \left( \frac{3-2p}{-1+2p} \right)}{\Gamma \left( \frac{1}{-1+2p} \right)}~.
\end{equation}
Thus, for $r \rightarrow \infty$ the electric potential behaves as:
\begin{equation}\label{asymp}
A_{t}(r \rightarrow \infty) \approx Q\frac{1-2p}{3-2p} r^{\frac{3-2p}{1-2p}}+ \ldots~,
\end{equation}
where $\ldots$ denotes terms that goes to zero at large distances. Eq. \eqref{asymp} shows that the field $A_{t}$ tends to zero at infinity for $1/2<p<3/2$ and diverges at infinity for $p<1/2$ and $p>3/2$. However, the above solution is not valid for $n=1/(2p-1)$ with $n>1$ being an integer. Fortunately, in that case we obtain analytical solutions that we present in the Appendix.
We can also determine the metric function $f(r)$ replacing \eqref{eqA} in \eqref{metricfunction}. We find
\begin{equation}\label{sol1}
\begin{aligned}
&f(r)=-\frac{\Lambda}{3}r^2-\frac{1}{3}\nu(6\alpha_{2}^{\prime}+\Lambda)r+k-\alpha_{2}^{\prime} \nu^2-2\alpha_{2}^{\prime} r(r+\nu) \ln \left(\frac{r}{r+\nu}\right)\\
&\quad+\frac{2\alpha_1 \eta 2^{p}p(2p-1) \nu^{(1-4p)/(1-2p)} }{p-1}r(r+\nu)\int_\infty^r\frac{r^{2p/(1-2p)}}{(r+\nu)^2} \,_2F_1\left[\frac{1}{2p-1},\,\frac{2(p-1)}{2p-1},\,\frac{4p-3}{2p-1}~,\,-\frac{r}{\nu}\right]dr~,
\end{aligned}
\end{equation}
where 
\begin{equation}
\alpha_2^{\prime}=\alpha_{2}-4 \alpha_1 \eta 2^{p} p \frac{\Gamma \left( \frac{2-2p}{1-2p} \right) \Gamma \left( \frac{3-2p}{-1+2p} \right)}{\Gamma \left( \frac{1}{-1+2p} \right)}~,
\end{equation}
and the following redefinition of the integration constant $Q$ have been taken into account: 
\begin{equation}
(Q^{2})^{p} = \frac{\alpha_1}{(\nu^2)^{2p/1-2p}}~,
\end{equation}
and the scalar field potential can be written as
\begin{equation}
\begin{aligned}
V(\phi)&=\frac{\Lambda}{3}\left(2+\cosh(\sqrt{2}\phi)\right)+2\alpha_2^{\prime}\left(-\sqrt{2}\phi\left[2
+\cosh(\sqrt{2}\phi)\right]+3\sinh(\sqrt{2}\phi)\right)\\
&\quad-2\alpha_1\eta p 2^p \frac{2p-1}{p-1} \{ 12 \left( e^{\sqrt{2}\phi}-1\right)^{2(1-p)/(2p-1)} \sinh^2(\phi/\sqrt{2}) \sinh (\sqrt{2} \phi) \\
& \times \,_2F_1\left[\frac{1}{2p-1},\,\frac{2(p-1)}{2p-1}~,\,\frac{4p-3}{2p-1},\,\frac{1}{1-e^{\sqrt{2}\phi}}\right] + \left(2+\cosh(\sqrt{2}\phi)\right)F(\phi)\} \\
& +\alpha_1 \eta 2^p \left( \frac{1}{4 \sinh^2 (\phi/\sqrt{2})}\right)^{2p/(1-2p)},
\end{aligned}
\end{equation}
where we have defined
\begin{equation}
\begin{aligned}
F(\phi)&=\nu^{(1-4p)/(1-2p)}\int_\infty^{\frac{\nu}{-1+e^{\sqrt{2}\phi}}}\frac{r^{2p/(1-2p)}}{(r+\nu)^2}\,_2F_1\left[\frac{1}{2p-1},\,\frac{2(p-1)}{2p-1},\,\frac{4p-3}{2p-1}~,\,-\frac{r}{\nu}\right]dr \\
& \quad = \int_\infty^{\frac{1}{-1+e^{\sqrt{2}\phi}}}\frac{r^{2p/(1-2p)}}{(r+1)^2}\,_2F_1\left[\frac{1}{2p-1},\,\frac{2(p-1)}{2p-1},\,\frac{4p-3}{2p-1}~,\,-r\right]dr.
\end{aligned}
\end{equation}
The integral above apparently depends on $\nu$; however, it can be shown numerically that the integral does not depend on $\nu$; so, in the last line of the above expression we have set $\nu=1$ for simplicity. Therefore, the potential $V(\phi)$ does not depend on the parameter $\nu$. 
We have used the reverse type procedure to obtain the potential, chosen in a way that the solution found from the remaining equations of motion with a metric ansatz and an electric field also solves the Klein-Gordon equation.
Then, we can investigate whether our system has a charged hairy black
hole solution. In Fig. \ref{plots0}  we plot the behavior of
the metric function $f\left( r\right) $ 
and the potential $V\left( \phi\right)$ in Fig. \ref{plots11},
for a choice of parameters  $\nu =0.5$, $\Lambda=-1$, $\alpha_1=1$, $\alpha_2=1$, $p=2,3,5$ and $k=\pm 1,0$. Also, in Fig. \ref{plott10} we plot the potential for other choices of the parameters. The metric function $f(r)$ changes
sign for low values of $r$, signaling the presence of a horizon,
while the potential tends to
$\Lambda=V(0)$ ($V(0)<0$)
as can be seen in Figs.
\ref{plots11} and \ref{plott10}. We observe a different behavior of the potential, while it is bounded from below in Fig. \ref{plott10}, this does not occur in Fig. {\ref{plots11}}. It is worth to mention that potentials with similar behavior have been considered for instance in \cite{Anabalon:2012ih, Gonzalez:2013aca, Gonzalez:2014tga, Fan:2015oca}. However, note that in Fig. \ref{plott10} the shift of the minimum of the potential depends on the parameter $p$.
Additionally, in Fig. \ref{plots00} we plot the metric function $f(r)$ for other choices of the parameters, and we observe that for certain values of the parameters the metric can describe a black hole with two horizons, $r_{+}$ and $r_{-}$. Moreover, the metric can describe an extremal black hole with degenerate horizons when $r_{+}=r_{-}$. Note that the full range of the $r$ coordinate covers all the values of $a(r)$ from the center $r=r_c$, where $a(r_{c})=0$, to infinity \cite{Bronnikov:2001tv}. So, for $\nu$ positive, the $r$ coordinate is in the range $0<r< \infty$, while for $\nu$ negative the range is $-\nu <r < \infty$.
 Finally, we have checked
the behavior of the Kretschmann scalar
$R_{\mu\nu\rho\sigma}R^{\mu\nu\rho\sigma}(r)$. Fig. \ref{figuraRR} shows that the Kretschmann scalar (with $\nu=0.5$) is singular at the center $r_{c}=0$, which is an essential property of these solutions, and the singularity is located in the dynamic region ($f(r_{c})<0$) for the numerical values considered (see Fig. \ref{plots0}). Furthermore, there is no
curvature singularity outside the horizon; therefore, the metric (\ref{sol1}) can describe a charged hairy black hole solution for certain values of the parameters.




Eq. \eqref{sol1} has an analytical solution for $p=3/4$ given by
\begin{equation}
\begin{aligned}
f(r)&=-\frac{\Lambda}{3}r^2-\frac{1}{3}\nu(6\alpha_{2}+\Lambda)r+k-\alpha_{2} \nu^2-2\alpha_{2}r(r+\nu) \ln \left(\frac{r}{r+\nu}\right)\\
&+3 \eta \alpha_1 2^{-1/4} \frac{\left(\nu(2r+\nu)+2r(r+\nu) \ln{\frac{r}{r+\nu}}\right)^2}{r (r+\nu)} ~.
\end{aligned}
\end{equation}
The electric potential is
\begin{equation}
A_{t}(r)=-\frac{Q}{\nu^2}\frac{2r+\nu}{r(r+\nu)}-\frac{2Q}{\nu^3}\ln{\frac{r}{r+\nu}}~,
\end{equation}
which goes to zero at infinity, and the self-interacting potential of the scalar field is

\begin{equation}
\begin{aligned}
V(\phi)&=\frac{\Lambda}{3}\left(2+\cosh(\sqrt{2}\phi)\right)+2\alpha_2\left(-\sqrt{2}\phi\left[2
+\cosh(\sqrt{2}\phi)\right]+3\sinh(\sqrt{2}\phi)\right)\\
&\quad+64 \eta 2^{3/4} \alpha_{1} \sinh^6(\phi/\sqrt{2})-3 \eta  \alpha_{1} 2^{-1/4}   (-28+16 \phi^2+(31+8 \phi^2) \cosh (\sqrt{2} \phi)\\
& \quad -4\cosh(2 \sqrt{2} \phi)+\cosh(3 \sqrt{2} \phi)-24 \sqrt{2} \phi \sinh(\sqrt{2} \phi))~.
\end{aligned}
\end{equation}

This solution is a special case of more general analytical solutions that we show in the Appendix. In Fig. \ref{plot} we show the behavior of $f(r)$ for $p=3/4$ and different values of $k$.

\begin{figure}[h]
\begin{center}
\includegraphics[width=0.4\textwidth]{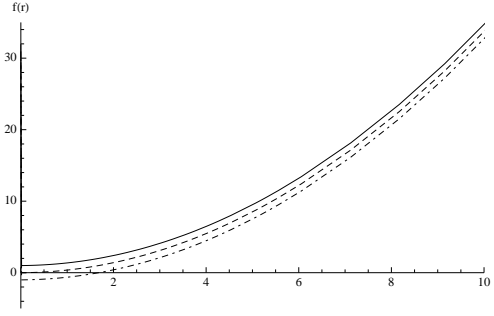}
\end{center}
\caption{The behavior of $f(r)$ for $\nu =1$, $\Lambda=-1$, $\eta=1/4$, $\alpha_1=1.5$, $\alpha_2=0.01$, $p=3/4$, $k=-1$ (dot-dashed line), $k=0$ (dashed line) and $k=1$ (continuous line).}
\label{plot}
\end{figure}

\subsection{Born-Infeld type electrodynamics}

\begin{figure}[h]
\begin{center}
\includegraphics[width=0.6\textwidth]{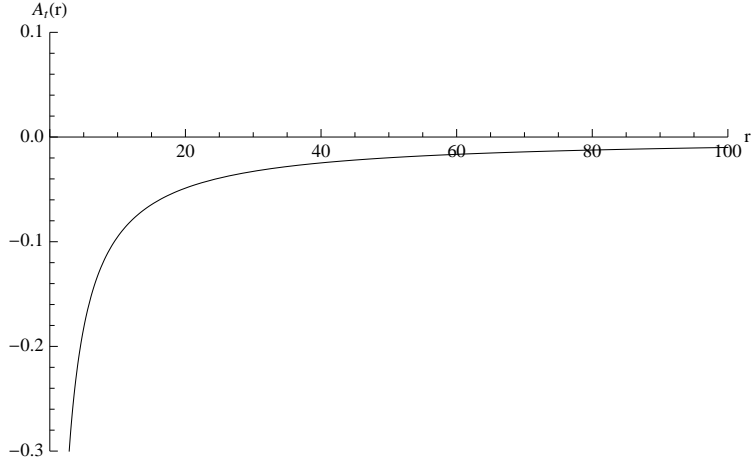}
\end{center}
\caption{The behavior of $A_{t}(r)$,  for $b=1$, $\nu=1$ and $\tilde{Q}=1$.} \label{scalar}
\end{figure}

In this section we consider a Born-Infeld type Lagrangian given by
\begin{equation}\label{BI}
\mathcal{L}(F^2)=4 b^2 \left( 1-\sqrt{1+\frac{F^2}{2b^2}} \right),
\end{equation}
where $b$ is the Born-Infeld coupling. In the limit $b \rightarrow \infty$ the Maxwell electrodynamics are recovered and in the limit $b \rightarrow 0$ this Lagrangian vanishes.
By inserting \eqref{BI} in (\ref{maxw}), we obtain straightforwardly the electric field
\begin{equation}
A_t^{\prime} (r)=\frac{\tilde{Q}}{\sqrt{r^2(r+\nu)^2+\frac{\tilde{Q}^2}{b^2}}}~.
\end{equation}
Then, by performing the change of variable $u=r(r+\nu)+\nu^2/12$, the scalar potential reads
\begin{equation}
A_t=\tilde{Q}\int \frac{du}{\sqrt{4u^3-g_{2}u-g_{3}}}~,
\end{equation}
where
\begin{equation}\label{wi}
g_{2}=\frac{1}{12} \left( \nu^4-48 \frac{\tilde{Q}^2}{ b^2} \right), \,\, g_{3}=-\frac{\nu^2}{216}\left( 144 \frac{\tilde{Q}^2}{ b^2}+\nu^4  \right).
\end{equation}
Therefore, the solution for the scalar potential is
\begin{equation}
A_{t}(r)=\tilde{Q} \wp^{-1}\left( r(r+\nu)+\nu^2/12;g_2,g_3 \right)~,
\end{equation}
where $\wp$ denotes the $\wp$-Weierstrass elliptic function, with the Weierstrass invariants $g_2$ and $g_3$ given in \eqref{wi}. In Fig. \ref{scalar} we plot $A_{t}(r)$ for $b=1$, $\nu=1$ and $\tilde{Q}=1$.

The metric function $f(r)$ reads
\begin{equation}\label{sol2}
\begin{aligned}
f(r)&=-\frac{\Lambda}{3}r^2-\frac{1}{3}\nu(6\alpha_{2}+\Lambda)r+k-\alpha_{2} \nu^2-2\alpha_{2}r(r+\nu) \ln \left(\frac{r}{r+\nu}\right) \\
&\quad +8\nu^4 \alpha_{1}^{2} r(r+\nu) \int_{\infty}^r\frac{\wp^{-1}\left( r(r+\nu)+\nu^2/12;g_2,g_3 \right)}{r^2(r+\nu)^2}dr~,
\end{aligned}
\end{equation}
where we have taken into account the following redefinition:
\begin{equation}
\tilde{Q}= \alpha_1 \nu^2~,
\end{equation}
and the potential can be written as 
\begin{equation}
\begin{aligned}
V(\phi)&=\frac{\Lambda}{3}\left(2+\cosh(\sqrt{2}\phi)\right)+2\alpha_2\left(-\sqrt{2}\phi\left[2
+\cosh(\sqrt{2}\phi)\right]+3\sinh(\sqrt{2}\phi)\right)\\
&\quad+4 b^2 \left(1-\frac{b e^{\sqrt{2}\phi}}{\sqrt{\alpha_{1}^2 (e^{\sqrt{2}\phi}-1)^4+b^2 e^{2\sqrt{2}\phi}}} \right)-64 \alpha_{1}^2\frac{\sinh^2(\phi/\sqrt{2})}{\sqrt{\text{csch}^4 (\phi/\sqrt{2})+16\alpha_{1}^2/b^2}}\\
&\quad -8\alpha_{1}^2 \left(2+\cosh(\sqrt{2}\phi)\right) F(\phi)-32 \alpha_{1}^2 \sinh^2(\phi/\sqrt{2}) \sinh(\sqrt{2}\phi)G(\phi)~,
\end{aligned}
\end{equation}

where we have defined
\begin{equation}\label{nonu}
\begin{aligned}
F(\phi)&=\nu^4 \int_{\infty}^{\frac{\nu}{e^{\sqrt{2}\phi}-1}}\frac{\wp^{-1}\left(r(r+\nu)+\nu^2/12;g_2,g_3 \right)}{r^2(r+\nu)^2}dr=\int_{\infty}^{\frac{1}{e^{\sqrt{2}\phi}-1}}\frac{\wp^{-1}\left(r(r+1)+1/12;g_2,g_3 \right)}{r^2(r+1)^2}dr~, \\
\quad G(\phi)&=\nu \wp^{-1} \left(\frac{\nu^2}{4}( \text{csch}^2 (\phi/\sqrt{2})+1/3);g_2,g_3\right)= \wp^{-1} \left(\frac{1}{4}(\text{csch}^2 (\phi/\sqrt{2})+1/3);g_2,g_3\right)~.
\end{aligned}
\end{equation}
Expression \eqref{nonu} apparently depends on the parameter $\nu$; however, it can be shown numerically that these expressions do not depend on $\nu$; for this reason, we have set $\nu=1$ in \eqref{nonu} for simplicity. Therefore, the potential $V(\phi)$ does not depend on the parameter $\nu$. 

\begin{figure}[h]
\begin{center}
\includegraphics[width=0.4\textwidth]{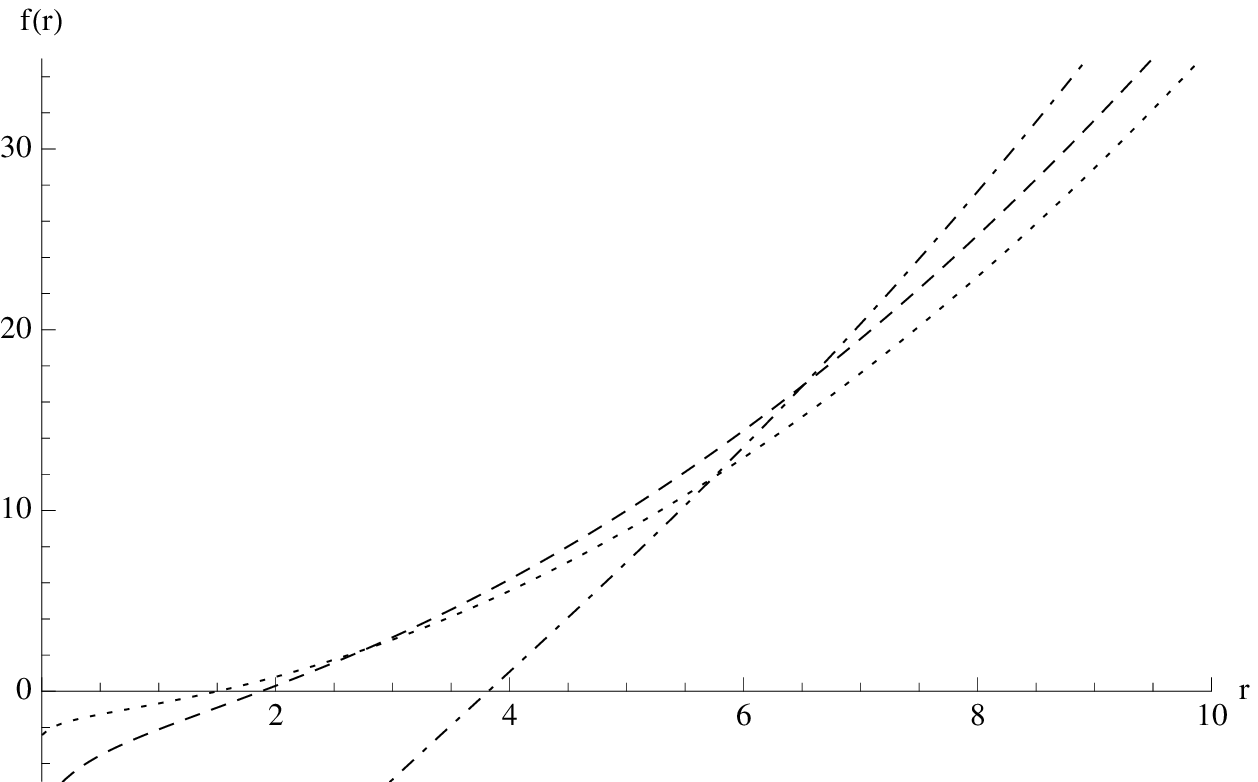}
\includegraphics[width=0.4\textwidth]{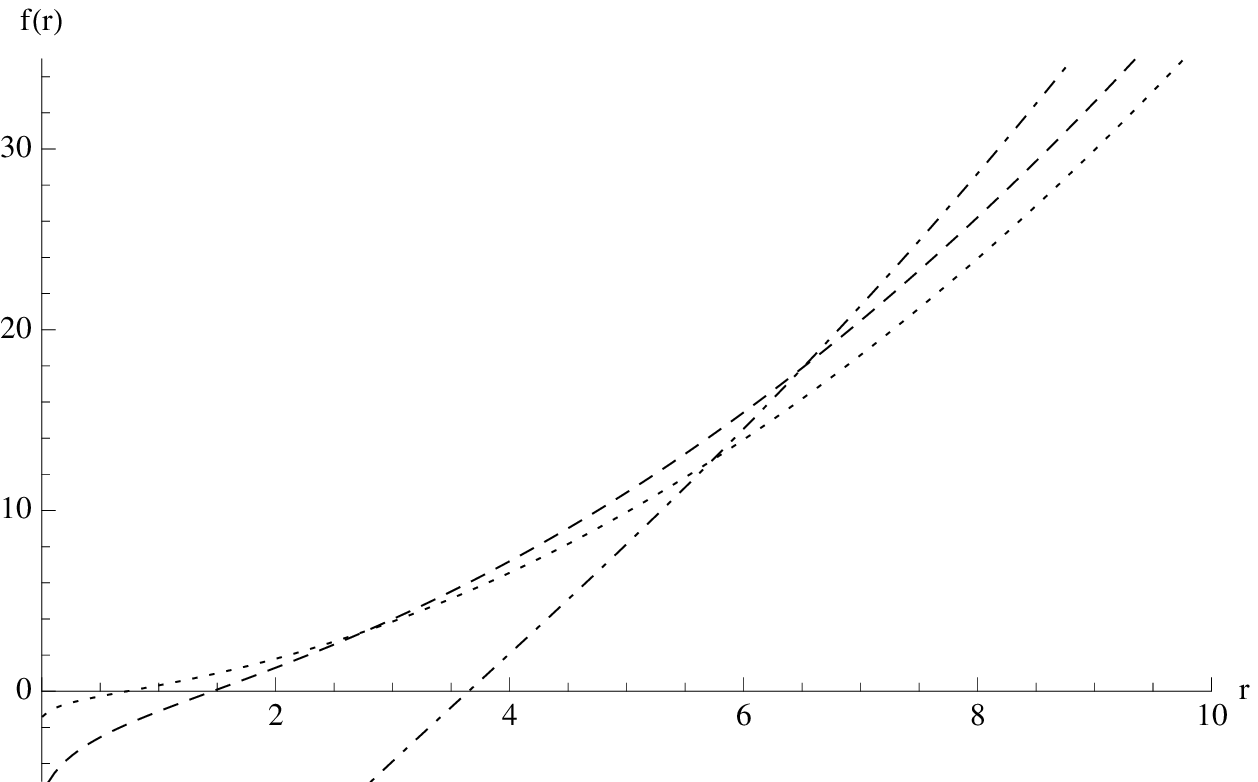}
\includegraphics[width=0.4\textwidth]{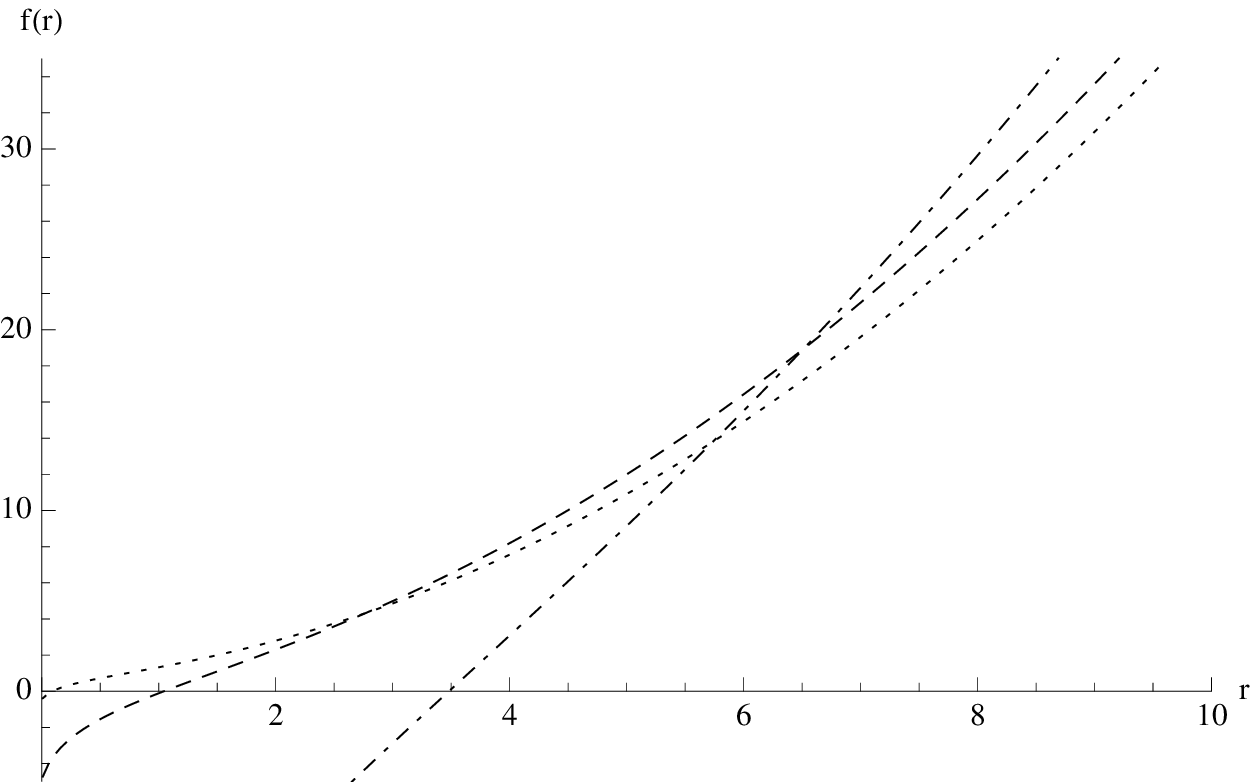}
\end{center}
\caption{The behavior of $f(r)$ for $\Lambda=-1$, $\alpha_1=0.01$, $\alpha_2=1.5$, $b=1$, $\nu=1$ (dotted line), $\nu=2$ (dashed line) and $\nu=5$ (dot-dashed line). Left figure for $k=-1$, right figure for $k=0$, and bottom figure for $k=1$.
}
\label{plots0BI}
\end{figure}

\begin{figure}[h]
\begin{center}
\includegraphics[width=0.4\textwidth]{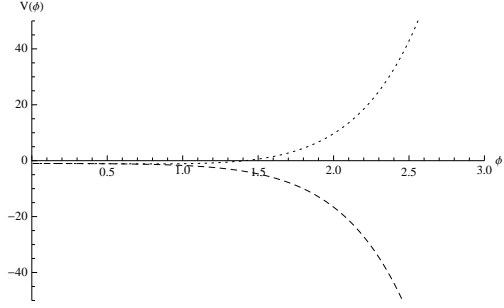}
\end{center}
\caption{The behavior of $V(\phi)$ for $\Lambda=-1$, $b=1$, $\alpha_1=0.01$, $\alpha_2=1.5$ (dashed line) and $\alpha_2=-1.5$ (dotted line).} \label{plots11BI}
\end{figure}

\begin{figure}[h]
\begin{center}
\includegraphics[width=0.4\textwidth]{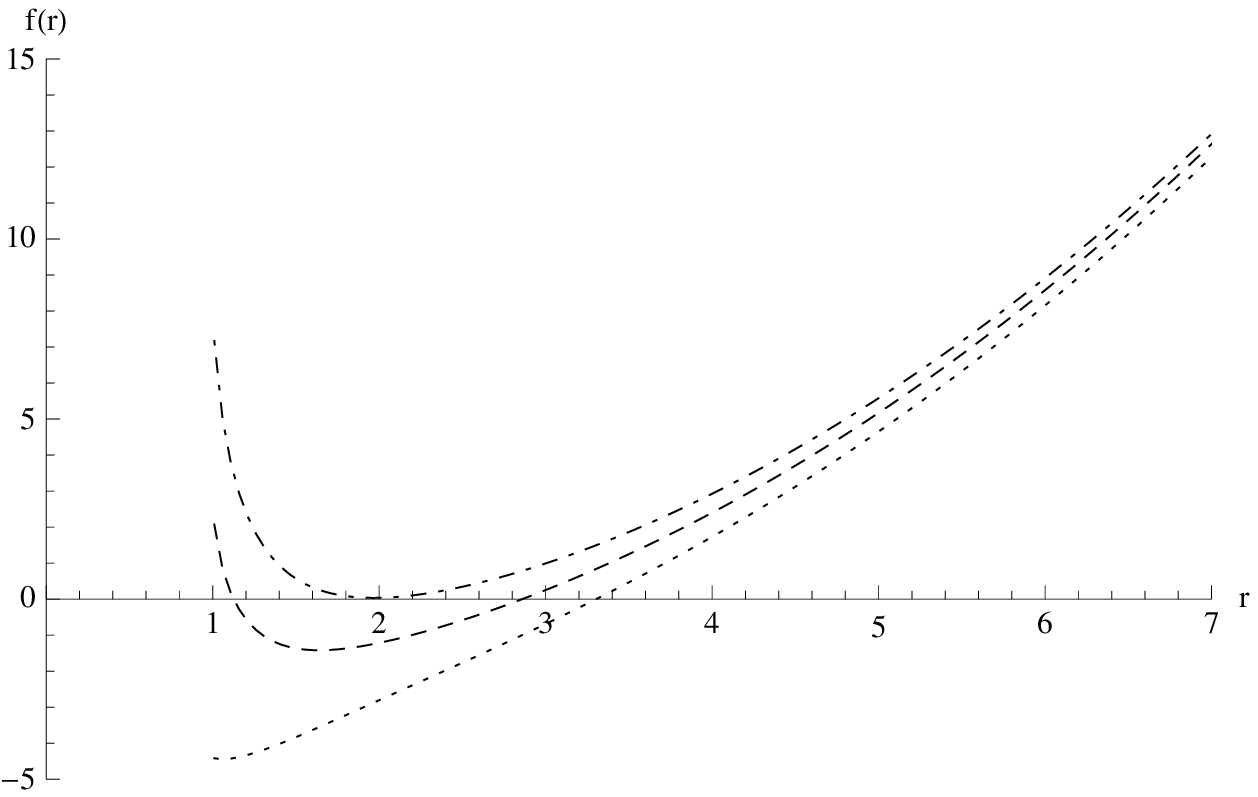}
\includegraphics[width=0.4\textwidth]{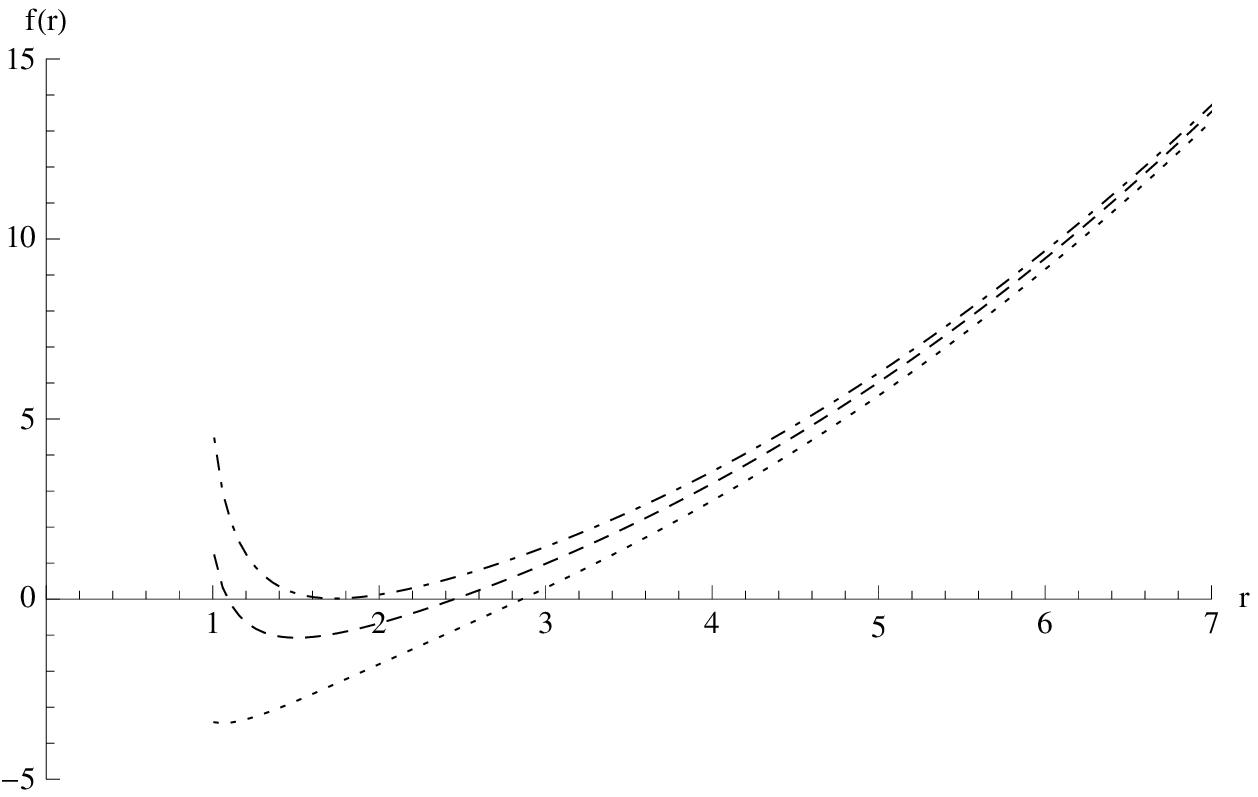}
\includegraphics[width=0.4\textwidth]{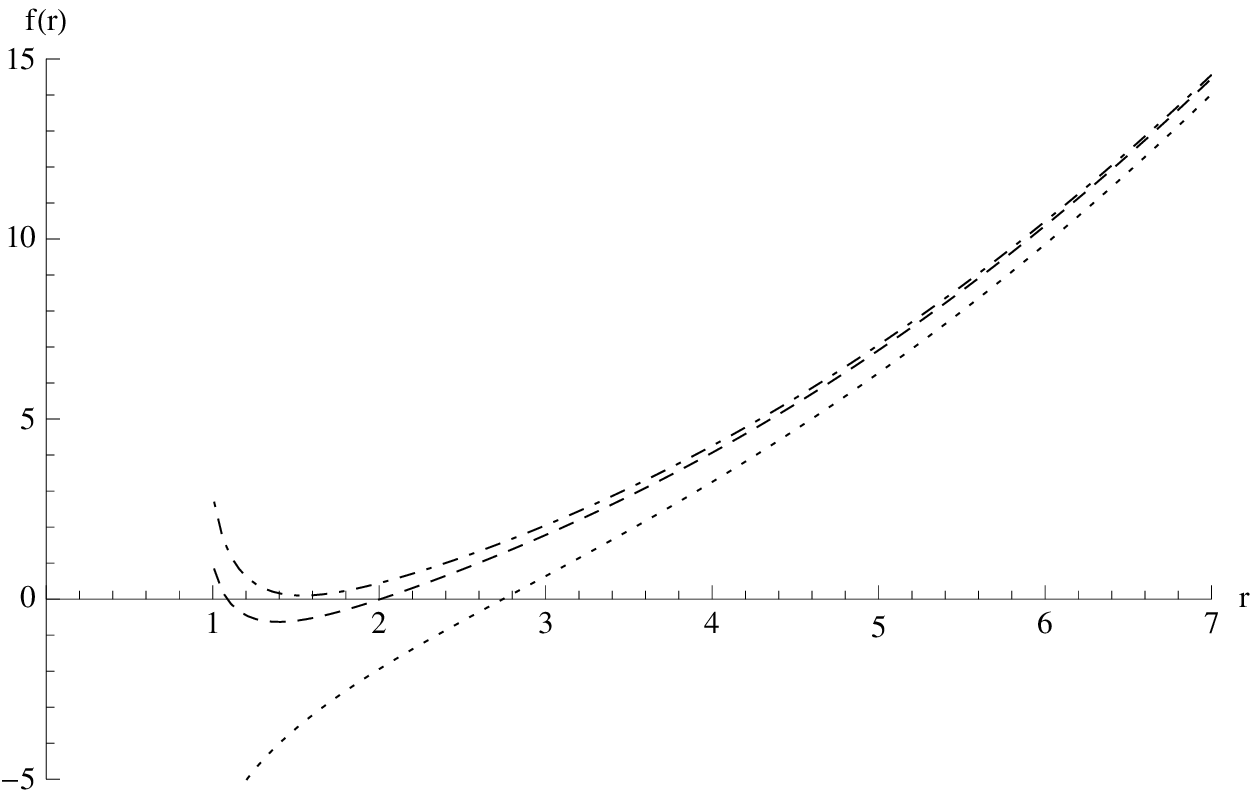}
\end{center}
\caption{The behavior of $f(r)$ for $\Lambda=-1$, $\alpha_1=1$, $\nu=-1$, $b=1$, (dotted line), $\nu=2$ (dashed line) and $\nu=5$ (dot-dashed line). Left figure for $k=-1$ and $\alpha_2=-15$ (dotted line), $\alpha_2=-8$ (dashed line), $\alpha_2=-2.5$ (dot-dashed line). Right figure for $k=0$ and $\alpha_2=-15$ (dotted line), $\alpha_2=-10$ (dashed line), $\alpha_2=-6.5$ (dot-dashed line). Bottom figure for $k=1$ and $\alpha_2=-20$ (dotted line), $\alpha_2=-11.5$ (dashed line), $\alpha_2=-9.5$ (dot-dashed line).
}
\label{plots1BI}
\end{figure}

\begin{figure}[h]
\begin{center}
\includegraphics[width=0.4\textwidth]{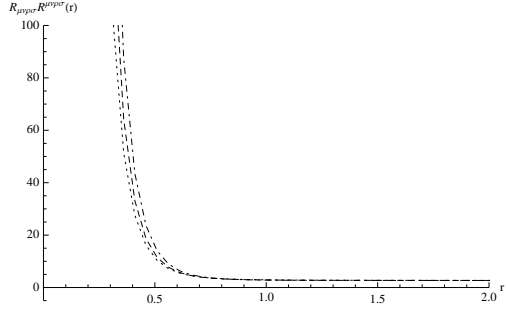}
\end{center}
\caption{The behavior of Kretschmann scalar $R_{\mu\nu\rho\sigma}R^{\mu\nu\rho\sigma}(r)$ as a function of $r$  for $\nu =0.5$, $\Lambda=-1$, $\alpha_1=1$, $\alpha_2=1$, $b=1$, $ k=-1$ (dotted line), $ k=0$ (dashed line) and $k=1 $ (dot-dashed line). 
} \label{figuraRRBI}
\end{figure}

Then, we can investigate whether our system has a charged hairy black
hole solution. In Fig. \ref{plots0BI}  we plot the behavior of
the metric function $f\left( r\right) $ 
and the potential $V\left( \phi\right)$ in Fig. \ref{plots11BI},
for a choice of parameters   $\Lambda=-1$, $\alpha_1=0.01$, $\alpha_2=1.5$, $b=1$ and $\nu = 1, 2, 5$. 
The metric function $f(r)$ changes
sign for low values of $r$ signaling the presence of a horizon,
while the potential tends to
$\Lambda=V(0)$ ($V(0)<0$),
as can be seen in Fig.
\ref{plots11BI}.
Additionally, in Fig. \ref{plots1BI} we plot the metric function $f(r)$ for other choices of the parameters, and we observe that for certain values of the parameters the metric can describe a black hole with two horizons, $r_{+}$ and $r_{-}$. Moreover, the metric can describe an extremal black hole with degenerate horizons when $r_{+}=r_{-}$. It is worth mentioning that in Fig. \ref{plots1BI} we have set $\nu=-1$. As we mentioned in the previous section, for $\nu$ negative the full range of the $r$ coordinate is $-\nu <r < \infty$, and $r_{c}=-\nu$ corresponds to the center. It can be shown that the scalar field and the Kretschmann scalar both diverge at the center $r_{c}=1$, which is an essential property of these solutions, and the singularity can be located either in a static region ($f(r_{c})>0$) as the dot-dashed and dashed lines illustrate in Fig. \ref{plots1BI} or in the dynamic region ($f(r_{c})<0$) as the dotted lines show in Fig. \ref{plots1BI}. Additionally, in Fig. \ref{figuraRRBI} we have plotted the behavior of the Kretschmann scalar $R_{\mu\nu\rho\sigma}R^{\mu\nu\rho\sigma}(r)$ for $\nu=0.5$, and it is shown that it is singular at the center $r_{c}=0$. The numerical results also show that there is no curvature singularity outside the horizon; therefore, the metric (\ref{sol2}) can describe a charged hairy black hole solution for certain values of the parameters.



\newpage

\section{Thermodynamics}
\label{secs.4}


In this section we will study the thermodynamics of the
 hairy black hole solutions found.
To compute the conserved charges we will apply the Euclidean formalism, we will work in the $\rho
=\sqrt{r\left( r+\nu \right) }$ coordinate, in which the metric
(\ref{metricBH}) can be written in the following  form:
\begin{equation}\label{metrica}
ds^{2}=-N(\rho)^2g(\rho)^2dt^2+\frac{1}{g(\rho)^2}d\rho^2+\rho^2d\Omega^2~,
\end{equation}
where
\begin{eqnarray}
N\left( \rho\right)^2 =\frac{\rho ^{2}}{\left( \frac{\nu^2}{4}+\rho^2\right) }~,\,\,
g\left( \rho\right)^{2} = \frac{f(\rho)}{\rho^2}\left( \frac{\nu^2}{4}+\rho^2\right)~,\,\,
\end{eqnarray}
and where $f(\rho)$ corresponds to the metric function $f(r)$ evaluated at $r=-\frac{\nu}{2}+\sqrt{\frac{\nu^2}{4}+\rho^2}$.
In these coordinates the scalar field is given by
\begin{equation}
\phi(\rho)=\frac{1}{\sqrt{2}}\ln\left[\frac{\nu+\sqrt{\nu^2+4\rho^2}}{-\nu+\sqrt{\nu^2+4\rho^2}}\right]~.
\end{equation}
Now, we go to the Euclidean time $t \rightarrow i \tau$, and for the Euclidean metric associated with   \eqref{metrica}, Maxwell equation \eqref{gmaxw} yields
\begin{equation}\label{mxrho}
\frac{A_t^{\prime}(\rho)}{N(\rho)}\frac{d\mathcal{L}(F^2)}{dF^2}=-\frac{\tilde{Q}}{\rho^2}~,
\end{equation}
where here and in the following the prime $^\prime$ denotes a derivative with respect to $\rho$. So, to apply the Euclidean formalism we consider the following  action:
\begin{equation}\label{EA}
I_E=-\frac{\beta\Omega}{4\pi} \int_{\rho+}^\infty\left(N(\rho)\mathcal{H}(\rho)+A_t\mathcal{P}^\prime\right)d\rho+B~,
\end{equation}
where 
\begin{equation}
\mathcal{H}(\rho)=\frac{\rho^2}{2G}\left[\frac{g(\rho)^2-1+\rho(g(\rho)^2)^\prime}{\rho^2}+\left(\frac{1}{2}g(\rho)^2\phi^{\prime 2}+V(\phi)\right)
-\left( \mathcal{L}(F^2)-\frac{4A_{t}^{\prime}(\rho)}{N(\rho)}\frac{\tilde{Q}}{\rho^2}\right)\right]
\end{equation}
is the reduced Hamiltonian which satisfies the constraint  $\mathcal{H}(\rho)=0$ and
\begin{equation}
\mathcal{P}(\rho)=-\frac{4A_{t}^{\prime}(\rho) \rho^2}{N(\rho)} \frac{d\mathcal{L}(F^2)}{dF^2}~,
\end{equation}
which satisfies $\mathcal{P}^\prime(\rho)=0$. By using Maxwell equation \eqref{mxrho} we can show that $\mathcal{P}(\rho)=4\tilde{Q}$ is a constant. Furthermore, $B$ is a surface term, $\beta=1/T$ is the period of
Euclidean time and finally $\Omega $ is the area of the spatial
2 section. We now compute the action when the field equations
hold. The condition that the permitted geometries should
not have conical singularities at the event horizon $\rho_+$ imposes
\begin{equation}\label{T}
T= \frac{N(\rho_+)\left(g(\rho_+)^{2}\right)^{\prime}}{4\pi }~.
\end{equation}
So, by using the grand canonical ensemble we can fix  the
temperature and the electric potential $\Phi =-A_{t}(\rho_+)$.
Then, the variation of the surface term yields
\begin{equation}
\delta B=\delta B_G+\delta B_\phi+\delta B_F~,
\end{equation}
where
\begin{equation}
\delta B_{G}=\beta \Omega N(\rho)\rho \delta g(\rho)^2 \Bigg\vert_{\rho+}^\infty~,
\end{equation}
\begin{equation}
\delta B_\phi=\beta \Omega N(\rho)\rho^2 g(\rho)^2 \phi^\prime \delta\phi\Bigg\vert_{\rho+}^\infty~,
\end{equation}
\begin{equation}
\delta B_F=\beta \Omega A_{t}(\rho) \delta\mathcal{P}\Bigg\vert_{\rho+}^\infty~.
\end{equation}

For the variation of the fields at large distances we must be careful with the integral appearing in the metric function \eqref{metricfunction}. It can be shown numerically that when $A_{t}$ tends to zero at infinity (for $1/2<p<3/2$ in power-law electrodynamics and in Born-Infeld type electrodynamics) the integral does not contribute to the conserved charges, and the variation of the fields at large distances yields 
\begin{equation}
\delta B_{G\infty}= \beta \Omega\left(-\frac{\Lambda \nu \rho}{6}-\alpha_2 \nu^2+\mathcal{O}\left(\frac{1}{\rho}\right)\right)\delta\nu~,
\end{equation}
\begin{equation}
\delta B_{\phi\infty}=\beta \Omega \left(\frac{\Lambda \nu \rho}{6}+\mathcal{O}\left(\frac{1}{\rho}\right)\right)\delta\nu~,
\end{equation}
\begin{equation}
\delta B_{F\infty}=\beta\Omega A_t(\infty)\delta\mathcal{P}~.
\end{equation}
On the other hand, when the electric potential diverges at infinity (for $p>3/2$ in power-law electrodynamics) the variation of the integral in the metric function \eqref{metricfunction} diverges too. However, it can be verified numerically that it cancels out exactly with the contribution coming from $\delta B_{F\infty}=\beta\Omega A_t(\infty)\delta\mathcal{P}$. So, we can have well defined conserved charges in that cases.

The variation of the fields at the horizon yields
\begin{equation}
\delta B_{G\rho_+}=-\frac{4\pi}{N(\rho)\beta}\delta\rho_+ ~,
\end{equation}
\begin{equation}
\delta B_{\phi\rho_+}=0~,
\end{equation}
\begin{equation}
\delta B_{F\rho_+}=\beta\Omega A_t(\rho_+)\delta\mathcal{P}~.
\end{equation}
Therefore
\begin{equation}
\delta B=-\frac{\beta\Omega}{3}\alpha_2\delta\nu^3+2\pi\Omega \delta\rho^2_+ +\beta\Omega\Phi\delta\mathcal{P}~.
\end{equation}
Thus, as the Euclidean action is related to the free energy $F$ through $I_E=-\beta F$ we obtain
\begin{equation}
\begin{aligned}
I_E&=S-\beta\mathcal{M}+\beta\Phi\mathcal{Q}\\
\end{aligned}~.
\end{equation}
Then, this relation makes it possible to  identify the mass ($\mathcal{M}$), the  entropy ($S$)  and the electric charge ($\mathcal{Q}$) as
\begin{equation}\label{M}
\mathcal{M}=\frac{\Omega}{3}\alpha_2\nu^3~,
S=\frac{\Omega}{4G}\rho^2_+~,
\mathcal{Q}=\Omega\mathcal{P}=4 \Omega \tilde{Q}~.
\end{equation}
Thus, for power-law electrodynamics the electric charge is given by
\begin{equation}
\mathcal{Q}_{(1)}=4 \Omega \eta 2^{p-1}p \alpha_1^{(2p-1)/2p}\nu^2~,
\end{equation}
and for the Born-Infeld type electrodynamics the electric charge is
\begin{equation}
\mathcal{Q}_{(2)}=4 \Omega \alpha_1 \nu^2~.
\end{equation}

\begin{figure}[h]
\begin{center}
\includegraphics[width=0.55\textwidth]{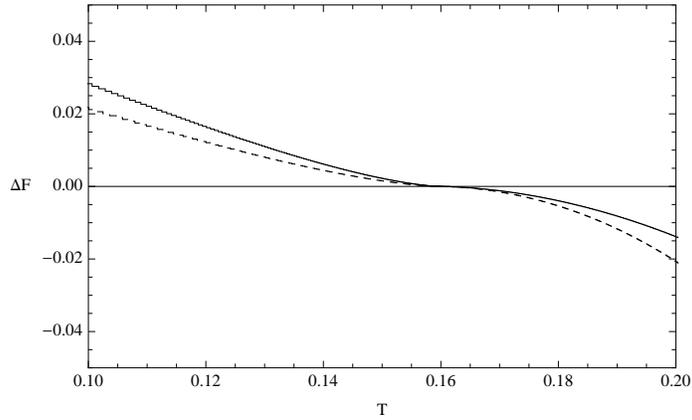}
\end{center}
\caption{The behavior of $\Delta F = F_{1}-F_{0}$ as a function of the temperature $T$ with $k=-1$, $\Omega=1$, $\Lambda=-3$, $\alpha_1=0.05$, $\alpha_2=-1$, $\eta=0.5$ with $p=0.9$ (continuous line) and $p=1.2$ (dashed line).
} 
\label{figura5}
\end{figure}

\begin{figure}[h]
\begin{center}
\includegraphics[width=0.55\textwidth]{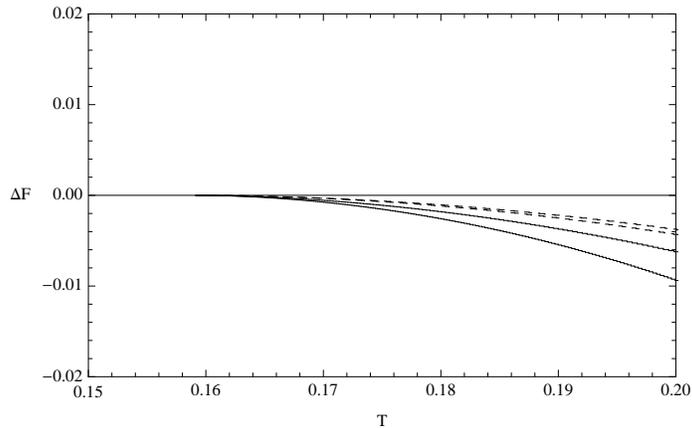}
\end{center}
\caption{The behavior of $\Delta F = F_{1}-F_{0}$ as a function of the temperature $T$ with $k=-1$, $\Omega=1$, $\Lambda=-3$, $\alpha_1=0.05$, $\alpha_2=-1$, $\eta=0.5$ with $p=2$ (continuous line) and $p=3$ (dashed line).
} 
\label{figura6}
\end{figure}

Having the temperature, mass, entropy and electric charge it is  possible to study
phase transitions between the nonlinearly charged black holes with scalar hair and  nonlinearly charged black holes without hair. For this analysis it is convenient to write the temperature of four-dimensional charged black holes with scalar hair (with $k=-1$) as
\begin{equation}\label{TS}
\begin{aligned}
T&=\frac{1}{4\pi}\{-\frac{1}{3}(2r_+ \Lambda + (12\alpha_2 + \Lambda)\nu)-2\alpha_2(2r_+ +\nu)\ln \left(\frac{r_+}{r_+ +\nu} \right)\\
\quad& +2^{2+p}\eta p(Q^{2})^{p}/Q \left( (2r_+ +\nu )\int_{\infty}^{r_+}\frac{A_t(r)}{r^2(r+\nu)^2}dr+r_+(r_+ +\nu)\frac{A_t(r_+)}{r_+^2(r_++\nu)^2} \right)\}~.
\end{aligned}
\end{equation}
On the other hand, in the absence of a scalar field, the field equations have as a solution topological nonlinearly charged black holes, which correspond to the metric presented in Eq. (15) of Ref. \cite{Zangeneh:2015wia} by setting the parameter $\alpha=0$ and $n=3$. The non-hairy black hole we consider here is also a generalization of the metric considered in \cite{Gonzalez:2009nn}, where a thermodynamic study of spherically symmetric black holes charged with power-law electrodynamics but with no cosmological constant was performed. 
The solution is
\begin{equation}
\begin{aligned}
ds^2 &=-\tilde{f}(\rho)dt^2+\frac{1}{\tilde{f}(\rho)}d\rho^2+\rho^2d\Omega^2~, \\
\tilde{f}(\rho) &=k-\frac{\Lambda}{3}\rho^2-\frac{m}{\rho}+\eta \frac{2^{p}(2p-1)^2(q^2)^p}{(3-2p)\rho^{\frac{2}{2p-1}}}~, \\
\tilde{A}_{t}(\rho)&=\frac{1-2p}{3-2p}q\rho^{\frac{3-2p}{1-2p}}~.
\end{aligned}
\end{equation}
The temperature,
entropy, mass and electric charge are given respectively by
\begin{equation}
\begin{aligned}
& T_{BH}=\frac{1}{4\pi}\left(-\frac{1}{\rho_+}-\Lambda \rho_+ + \eta 2^{p}(1-2p)(q^2)^p\rho_+^{\frac{1+2p}{1-2p}}\right)~,\\
\quad & S_{BH}=2\pi \Omega \rho _{+}^{2}~,\\
\quad & M_{BH}= -\Omega \rho_+\left(1+\frac{\rho_+^2\Lambda}{3}-\eta \frac{2^{p}(2p-1)^2\rho_+^{\frac{2}{1-2p}}(q^2)^p}{3-2p}\right)~,\\
\quad & Q_{BH}=2 \eta \frac{\Omega p 2^p (q^2)^p }{q}~,
\end{aligned}
\end{equation}
where
\begin{equation}
q=\frac{(2p-3)\Phi}{(1-2p)\rho_+^{\frac{3-2p}{1-2p}}}~.
\end{equation}
So, the horizon radius $\rho _{+}$
can be written as a
function of the temperature and of the electric potential. Now, in order to find phase transitions, 
we must consider both black holes in the same grand canonical ensemble, i.e., at the same $T$ and $\Phi$. Making $T$ and $\Phi$ equal for both black holes and by considering the free energy $F$
\begin{equation}
F =F(T,\Phi)=\mathcal{M}-TS-\Phi \mathcal{Q}~,
\end{equation}
we plot
the difference of the free energies for the nonlinearly charged black hole ($F_1$) and the nonlinearly charged black hole with scalar hair ($F_0$), $\Delta F= F_1-F_0$ as a function of the temperature in Figs.
\ref{figura5} and \ref{figura6}. In Fig. \ref{figura5} we show that there is a
second-order phase transition at the fixed critical temperature $T_c \approx 0.16$ for low values of $p$, and the topological nonlinearly charged hairy black hole dominates for temperatures lower than the critical, whereas for temperatures higher than the critical the topological nonlinearly charged black hole is thermodynamically preferred. In Fig. \ref{figura6} we show that there is not phase transitions for high values of $p$ and the charged black hole without scalar hair is thermodynamically preferred. It is worth mentioning that since $\nu$ and $\tilde{Q}$ are related, the temperature and the electric potential are not independent; thus, we can express the free energy as a function of $T$ or $\Phi$ only. This is similar to what happens in \cite{Martinez:2010ti} for a charged hairy black hole in a linear electrodynamic.

\section{Conclusions}
\label{secs.5}

We have considered a gravitating system consisting of a scalar
field minimally coupled to gravity with a self-interacting
potential and a U(1) nonlinear electromagnetic field. We solved the
coupled 
field equations 
with a profile of
the scalar field which falls sufficiently fast outside the black
hole horizon.  For a range of  values of the scalar field
parameter, which characterizes its behavior,  we found numerically charged
hairy black hole solutions in power-law and Born-Infeld type electrodynamics, with the scalar field  regular
everywhere outside and on the event horizon. Also, in the case of power-law electrodynamics, we found analytical hairy black hole solutions for some special values of the exponent $p$ in the range $1/2<p<1$.
Then, we studied the thermodynamics for both electrodynamics and phase transitions of our black
hole solutions in power-law electrodynamics, and we showed that there is a
second-order phase transition for low values of $p$. Moreover, at a low temperature, the topological nonlinearly charged hairy black hole is thermodynamically preferred, whereas the topological charged black hole without scalar hair is thermodynamically preferred at a high temperature for power-law electrodynamics. Interestingly enough, these phase transitions occur at a fixed critical temperature and do not depend on the exponent $p$ of the nonlinearity electrodynamics. On the other hand, we showed that there is not phase transitions for high values of $p$.
This picture is consistent with the findings of the
application of the AdS/CFT correspondence to condensed matter
systems. In these systems there is a critical temperature below
which the system undergoes a phase transition to a hairy black
hole configuration at a low temperature. This corresponds in the
boundary field theory to the formation of a condensation of the
scalar field.


\acknowledgments 

We would like to thank the anonymous referee for valuable comments which help us to improve the quality of our paper. This work was partially funded by Comisi\'{o}n Nacional de Ciencias y Tecnolog\'{i}a through FONDECYT grants 11140674 (PAG), by Direcci\'{o}n de Investigaci\'{o}n y Desarrollo de la Universidad de La Serena (Y.V.) and partially supported by grants from Comisi\'on Nacional de Investigaci\'on Cient\'ifica y Tecnol\'ogica CONICYT, Doctorado Nacional  2016 (21160784) (J.B.O.). P. A. G. and J. B. O. acknowledge the hospitality of the Universidad de La Serena, where part of this work was carried out. P.A.G. acknowledges the hospitality of the National Technical University of Athens.


\appendix
\section{Analytical solutions}

In this appendix we present some analytical hairy black hole solutions with power-law electrodynamics for some particular values of the exponent $p$. For $p=\frac{1}{2}+\frac{1}{2n}$, with $n>1$ being an integer number, so $\frac{1}{2}<p<1$, the electric field is given by 
\begin{equation}
A_{t}(r)=Q I1(r;n)~,
\end{equation}
where we have defined
\begin{equation}
\begin{aligned}
& I1(r;n)\equiv \int \frac{dr}{(r(r+\nu))^n}=\frac{\frac{\nu}{2}+r}{-\frac{\nu^2}{2}(n-1)(r(r+\nu))^{n-1}} \\
& \quad +\frac{\frac{\nu}{2}+r}{(n-1)} \sum_{j=1}^{n-2} \left( \prod_{i=1}^j \frac{2n-2i-1}{n-i-1} \right) \frac{1}{(-\frac{\nu^2}{2})^{j+1} (r(r+\nu))^{n-j-1}}  + \frac{1}{\nu \left(-\frac{\nu^2}{2}\right)^{n-1}} \prod_{i=1}^{n-1} \frac{2n-2i-1}{n-i} \ln \left(\frac{r}{r+\nu} \right).
\end{aligned}
\end{equation}
So, using this expression, we find that the metric function is given by
\begin{equation}
\begin{aligned}
f(r)&=-\frac{\Lambda}{3}r^2-\frac{1}{3}\nu(6\alpha_{2}+\Lambda)r+k-\alpha_{2} \nu^2-2\alpha_{2}r(r+\nu) \ln \left(\frac{r}{r+\nu}\right)\\
&\quad +2^{\frac{3}{2}+\frac{1}{2n}} \eta \frac{(n+1)}{n} Q^{\frac{(n+1)}{n} } \{ -r(r+\nu)\frac{1}{\frac{\nu^2}{2}(n-1)}\left(\frac{\nu}{2}I1(r;n+1)+I2(r;n+1) \right) \\
&\quad + r(r+\nu) \sum_{j=1}^{n-2} \left( \prod_{i=1}^{j} \frac{2n-2i-1}{n-i-1} \right) \frac{1}{ (n-1)\left( -\frac{\nu^2}{2} \right)^{j+1}} \left( \frac{\nu}{2} I1(r;n-j+1)+I2(r;n-j+1)\right) \\
&\quad - \frac{1}{\left( -\frac{\nu^2}{2}\right)^{n-1} \nu^4} \left( \prod_{i=1}^{n-1} \frac{2n-2i-1}{n-i} \right) \left(\nu+(r+\nu) \ln \left( \frac{r}{r+\nu}\right) \right) \left(\nu+r\ln\left( \frac{r}{r+\nu}\right) \right)\}~,
\end{aligned}
\end{equation}
where we have defined
\begin{equation}
\begin{aligned}
I2(r;n)&\equiv \int \frac{rdr}{(r(r+\nu))^n}= \frac{r}{\nu (n-1) (r(r+\nu))^{n-1}}+\frac{2n-3}{\nu(n-1)}I1(r;n-1).
\end{aligned}
\end{equation}

On the other hand, for $p=\frac{1}{2}+\frac{1}{2m-1}$, with $m>1$ being an integer number, so $\frac{1}{2}<p<1$, the electric field is given by 
\begin{equation}
A_{t}(r)=Q I3(r;m)~,
\end{equation}
where
\begin{equation}
\begin{aligned}
I3(r;m)&\equiv \int \frac{dr}{(r(r+\nu))^{\frac{2m-1}{2}}}=\frac{\frac{\nu}{2}+r}{-\frac{\nu^2}{2}\frac{2m-3}{2}(r(r+\nu))^{\frac{2m-3}{2}}} \\
& \quad +\frac{\frac{\nu}{2}+r}{\frac{2m-3}{2}} \sum_{j=1}^{m-1} \left( \prod_{i=1}^j \frac{2m-2i-2}{\frac{2m-1}{2}-i-1} \right) \frac{1}{(-\frac{\nu^2}{2})^{j+1} (r(r+\nu))^{\frac{2m-2j-3}{2}}}~,
\end{aligned}
\end{equation}
and the metric function is
\begin{equation}
\begin{aligned}
f(r)&=-\frac{\Lambda}{3}r^2-\frac{1}{3}\nu(6\alpha_{2}+\Lambda)r+k-\alpha_{2} \nu^2-2\alpha_{2}r(r+\nu) \ln \left(\frac{r}{r+\nu}\right)\\
&\quad +2^{\frac{3}{2}+\frac{1}{2m-1}} \eta \frac{2m+1}{2m-1} Q^{\frac{2m+1}{2m-1}} \{ -r(r+\nu)\frac{1}{\frac{\nu^2}{2}(2m-3)}\left(\frac{\nu}{2}I3(r;m+1)+I4(r;m+1) \right) \\
&\quad + r(r+\nu) \sum_{j=1}^{m-1} \left( \prod_{i=1}^{j} \frac{2m-2i-2}{\frac{2m-1}{2}-i-1} \right) \frac{1}{ (\frac{2m-3}{2})\left( -\frac{\nu^2}{2} \right)^{j+1}} \left( \frac{\nu}{2} I3(r;m-j+1)+I4(r;m-j+1)\right)\}~,
\end{aligned}
\end{equation}
where
\begin{equation}
\begin{aligned}
I4(r;m)&\equiv \int \frac{rdr}{(r(r+\nu))^{\frac{2m-1}{2}}}= \frac{r}{\frac{2m-3}{2} \nu (r(r+\nu))^{\frac{2m-3}{2}}}+\frac{2(m-2)}{\frac{2m-3}{2}\nu}I3(r;m-1)~.
\end{aligned}
\end{equation}

\end{document}